\documentclass[11pt, letterpaper]{article}
\usepackage[utf8]{inputenc}
\usepackage[margin=1.5cm]{geometry}
\usepackage{titlesec}
\usepackage{tabu}
\usepackage{enumitem}
\usepackage{amssymb}
\usepackage{xcolor}
\newlist{selectlist}{itemize}{2}
\setlist[selectlist]{label=$\square$,leftmargin=*,noitemsep,topsep=0pt}

\usepackage{lmodern}

\usepackage{hyperref}
\hypersetup{
    colorlinks=true,
    linkcolor=blue,
    filecolor=magenta,      
    urlcolor=blue,
}

\usepackage[AMcore,AMbiblio,AMref,AMmath,AMtikz,optEnglish,optBiber]{AMlatex}

\usepgfplotslibrary{groupplots}
\tikzset{external/system call={pdflatex \tikzexternalcheckshellescape -halt-on-error
		-interaction=batchmode -jobname "\image" "\texsource" --extra-mem-bot=100000000 --extra-mem-top=100000000 --pool-size=10000000 --buf-size=10000000}}

\makeatletter
\let\blx@rerun@biber\relax
\makeatother

\usepackage{graphicx}

\graphicspath{{figures/},{figures/history},{figures/hw_description},{figures/build_instructions},{figures/validation},{figures/operating}}
 
\usepackage{float}

\usepackage{colortbl}
\usepackage{multirow}

\usepackage{subcaption}

\usepackage{tcolorbox}
\newtcolorbox{NOTES}{parbox=false,
	boxrule=0pt,leftrule=3mm,rightrule=3mm,boxsep=0pt,arc=0pt,outer arc=0pt,
	left=3mm,right=3mm,top=3mm,bottom=3mm,toptitle=1mm,bottomtitle=1mm,oversize,
	colback=blue!5!white,colframe=blue}

\titleformat{\section}[block]{\hspace{1em}\bfseries}{\thesection.}{0.5em}{} 
\titleformat{\subsection}[block]{\hspace{1em}}{\thesubsection}{0.5em}{}
\titleformat{\subsubsection}[block]{\hspace{1em}}{\thesubsubsection}{0.5em}{}

\usepackage{listings}
\usepackage{accsupp}
\newcommand{\noncopynumber}[1]{
	\BeginAccSupp{method=escape,ActualText={}}
	#1
	\EndAccSupp{}
}
\definecolor{codeBG}{rgb}{0.9,0.9,0.9}
\lstdefinestyle{Python}{
	backgroundcolor = \color{codeBG},
	language = Python,
	basicstyle=\fontfamily{pcr}\scriptsize,
	breakatwhitespace=false,         
	breaklines=true,                 
	captionpos=b,                    
	keepspaces=true,                                 
	showspaces=false,                
	showstringspaces=false,
	showtabs=false,                  
	tabsize=2,
	inputencoding=latin1,
	numbers=left,
	numberstyle=\tiny\noncopynumber,
	columns=flexible,
	xleftmargin=6ex,
	emph=[1]{OasisBoard},emphstyle=[1]\color{TUMBlue}\bfseries,
	emph=[2]{connect,set_parameters,acquire,save_data_h5},emphstyle=[2]\color{TUMOrange}\bfseries,
}

\usepackage{eurosym}

\newcommand{\SpecTable}{\bgroup\def\arraystretch{1.15}\setlength\tabcolsep{0.6mm}\begin{tabular}{|l|l|}
		\hline Available channels & 8 IEPE channels with SMB connector \\\hline
		Sampling frequency & \SI{100}{\kilo\hertz} per channel @ $18$-bit resolution \\\hline
		Voltage ranges & $\pm\SI{2.5}{\volt}$, $\pm\SI{5}{\volt}$, $\pm\SI{6.25}{\volt}$, $\pm\SI{10}{\volt}$, $\pm\SI{12.5}{\volt}$ \\\hline
		Bill of materials cost & Approximately \textdollar$160$ / $140$\;\euro \\\hline
	\end{tabular}\egroup}

\usepackage{blindtext}

\usepackage{pifont}

\begin{document}

\setlength{\parindent}{0pt}
\setlength{\parskip}{10pt}

% !TeX spellcheck = en_US

\begin{flushleft}
%Insert title
\textbf{Article title}\\ OASIS-ERIS: Open Acquisition System for IEPE Sensors - Electrically Refined and Inter-device Synchronization

%Insert Authors
\textbf{Authors}\\ Oliver M. Zobel$^{*1}$, Johannes Maierhofer$^1$, Daniel J. Rixen$^1$

%Insert Affiliations
\textbf{Affiliations}\\ $^1$: Chair of Applied Mechanics, TUM School of Engineering and Design,
Technical University of Munich, Boltzmannstr. 15, 85748 Garching, Germany

%Insert Contact Email
\textbf{Corresponding author’s email address}\\ oliver.zobel@tum.de

%Insert Abstract
\textbf{Abstract}\\ 
\textit{OASIS-ERIS}, the Open Acquisition System for IEPE Sensors - Electrically Refined and Inter-device Synchronization, is the successor to the previously released \textit{OASIS-UROS}. It is based on the \textit{Analog Devices AD7606C-18} analog-to-digital converter, offering 18-bit resolution on eight simultaneously sampled channels, and the \textit{Espressif ESP32-S3} microcontroller. The laboratory usability of \textit{OASIS-ERIS} is significantly improved compared to \textit{OASIS-UROS} due to the added inter-device synchronization. By physically connecting the sample clocks across devices, a truly synchronized data acquisition can be achieved with an arbitrary number of acquisition boards. Additionally, a migration of the firmware to \textit{ESP-IDF} allows a major increase of the achievable sampling frequency to \SI{100}{\kilo\hertz}. This paper documents the hardware and software stack of \textit{OASIS-ERIS}, and provides all materials necessary for reproduction. Further, the performance of \textit{OASIS-ERIS} is tested against a commercial measurement system in an Experimental Modal Analysis campaign. The results show that \textit{OASIS-ERIS} closely matches the performance of the commercial system, making it a viable replacement for academia and projects with tightly constrained budgets.

%Insert Keywords

\textbf{Specifications table}\\
\vskip 0.2cm
\tabulinesep=1ex
\begin{tabu} to \linewidth {|X|X[3,l]|}
	\hline \textbf{Hardware name} & OASIS-ERIS \\
	\hline \textbf{Subject area} & Educational tools and open-source alternatives to existing commercial tools \\
	\hline \textbf{Hardware type} & Measuring physical properties and in-lab sensors \\ 
	\hline \textbf{Closest commercial analog} &
	OROS MODS OR10-DAQ-8 or similar compact acquisition systems with 8 IEPE channels and compact form factor\\
	\hline \textbf{Open-source license} & CC-BY 4.0 (Hardware) / MIT (Software)\\
	\hline \textbf{Cost of hardware} & $\approx$ \textdollar$160$ / $140$\;\euro \\
	\hline \textbf{Source file repository} & \url{https://doi.org/10.5281/zenodo.21426150}\\
	\hline \textbf{OSHWA certification UID} & DE000177 \\\hline
\end{tabu}
\end{flushleft}

\newpage

% !TeX spellcheck = en_US

\section{Hardware in context}

Today, the importance of open-source software is becoming increasingly apparent. On the one hand, there are huge successful projects like \textit{Linux}\footnote{\url{https://kernel.org/}} that heavily support or even carry our whole digital infrastructure; or projects like \textit{FFmpeg}\footnote{\url{https://ffmpeg.org/}}, which is the basis for many media transcoding programs and services. On the other hand, there are projects aimed directly at the scientific community, for example, the well-known \textit{NumPy}\footnote{\url{https://numpy.org/}}~\cite{numpy} and \textit{SciPy}\footnote{\url{https://scipy.org/}}~\cite{scipy} packages. All of those open-source projects benefit from the key advantages of open-source:\vspace{-0.4cm}
\begin{itemize}
	\item Openness: anyone can view and audit the source code.\vspace{-0.2cm}
	\item Freedom: anyone can freely copy, modify, and redistribute projects (when the license permits it).\vspace{-0.2cm}
	\item Flexibility and Customization: anyone can fully customize an existing project to fit their requirements.\vspace{-0.2cm}
	\item Community and Collaboration: anyone can contribute to existing projects and improve them for everyone.\vspace{-0.2cm}
	\item Ownership: a rightfully obtained copy cannot be taken away.
\end{itemize}\vspace{-0.4cm}
While open-source software is readily available today for structural dynamic applications (see, for instance, the review in~\cite{pythonsd}), only a few open-source hardware projects specifically tailored to structural dynamics exist. The benefits of open-source hardware are the same as for software, but an additional advantage exists: an often significantly lower cost compared to commercial solutions. This advantage of open-source hardware is also the key motivation for the \textit{Open Acquisition System for IEPE Sensors} (\textit{OASIS})\footnote{\url{https://gitlab.com/oasis-acquisition}} project.

An open acquisition system, which is significantly cheaper than commercial alternatives, is beneficial for academia in teaching and research. For teaching, cheap hardware enables, for instance, lab courses with hands-on experience for many students. For research, a low-cost system enables the permanent incorporation of a data acquisition system into test rigs or into measurements with a high risk of hardware destruction, e.g., on drones or race cars.

\begin{figure}[H]
	\centering
	\scriptsize
	\def\svgwidth{\textwidth}
	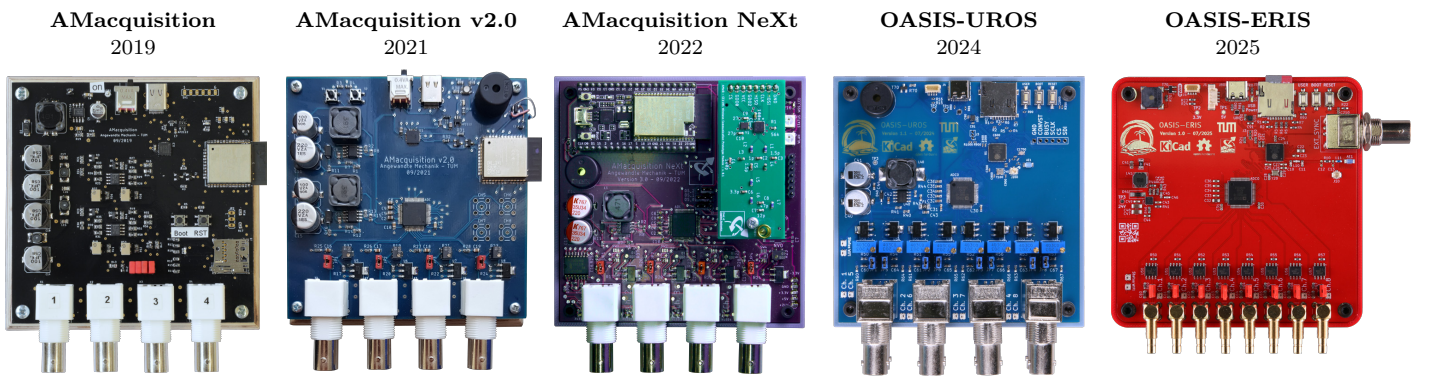
	\caption{Development history overview of OASIS boards in chronological order.}
	\label{fig:history}
\end{figure}

The open acquisition system described in this article, \textit{OASIS-ERIS} (\textit{Open Acquisition System for IEPE Sensors - Electrically Refined and Inter-device Synchronization}), is the latest version to be released. It follows its predecessors, as seen in \cref{fig:history}, which the authors also developed. After the first two internal development versions \textit{AMacquisition} and \textit{AMacquisition v2.0}, the original \textit{OASIS} board (internally called \textit{AMacquisition NeXt}) was released in 2022~\cite{OASISPaper}. Following this, a major overhaul and cleanup were performed, resulting in the release of \textit{OASIS-UROS} (\textit{Open acquisition system for IEPE sensors - upgraded, refined, and overhauled software}) in 2024~\cite{OASISUROS}. Most importantly, this release saw an extensive documentation of the device's architecture.

\vspace{-0.4cm}

Therefore, this article is written as an update to the original article, mainly detailing changes and new features. For a comparison of \textit{OASIS} against commercial systems, the reader is referred to the \textit{OASIS-UROS} article~\cite{OASISUROS}.
% !TeX spellcheck = en_US
\section{Hardware description}

\begin{figure}[H]
	\centering
	\scriptsize
	\def\svgwidth{\textwidth}
	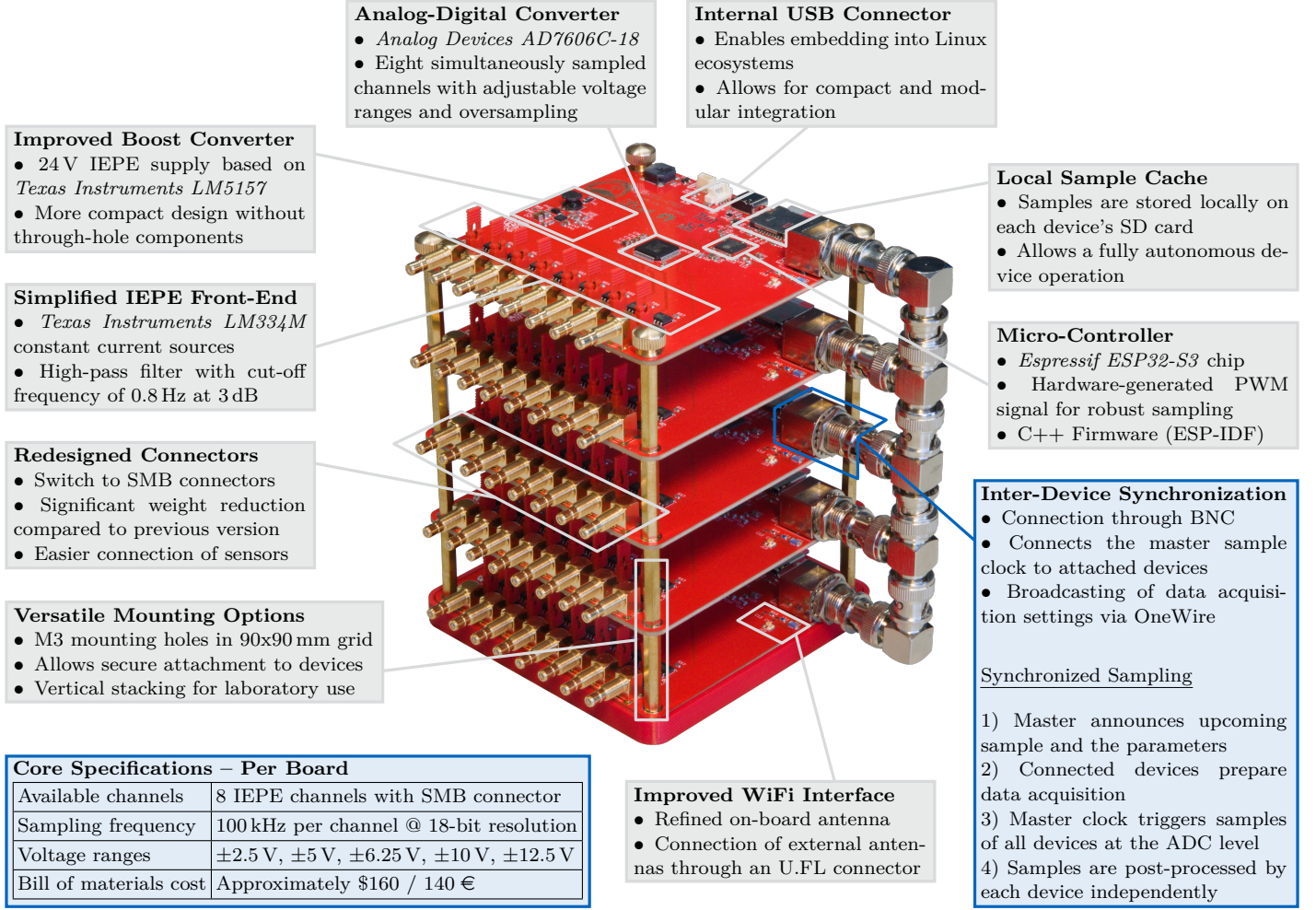
	\caption{Overview of \textit{OASIS-ERIS}'s key features. The figure shows five \textit{OASIS-ERIS} boards that are physically connected and synchronized through BNC connectors.}
	\label{fig:HWOverview}
\end{figure}

In this section, the features of the \textit{OASIS-ERIS} board are described. Compared to the previous \textit{OASIS-UROS}~\cite{OASISUROS}, the most notable new feature is the added inter-device synchronization. This feature allows synchronizing together as many \textit{OASIS-ERIS} as desired. An overview of \textit{OASIS-ERIS}'s key features is given in \cref{fig:HWOverview}. In the following, the hardware and software features, as well as the new inter-device synchronization, are described in detail.

\vspace{-0.4cm}

\subsection{Hardware features}

As \textit{OASIS-UROS}, \textit{OASIS-ERIS} is still based around the \textit{Analog Devices AD7606C-18} analog-to-digital converter (ADC) with an 18-bit resolution and eight simultaneously sampled channels. Through SPI (Serial Peripheral Interface), it is possible to configure each channel's input voltage range ($\pm\SI{2.5}{\volt}$, $\pm\SI{5}{\volt}$, $\pm\SI{6.25}{\volt}$, $\pm\SI{10}{\volt}$, $\pm\SI{12.5}{\volt}$) and enable oversampling to reduce noise~\cite{ADCDatasheet}.

The system is controlled by an \textit{Espressif ESP32-S3}~\cite{ESP32S3TechRef} micro-controller with two cores and WiFi capabilities. Using both cores of the \textit{ESP32-S3} enables fast data sampling, where one core acquires the samples from the \textit{AD7606C-18} ADC using Octal SPI and stores them in cache pages in RAM, while the other core writes the samples from the other cache page to the micro SD card; this is described in detail in the \textit{OASIS-UROS} publication~\cite{OASISUROS}.

Significant changes were implemented in the front-end design: instead of the large, heavy dual BNC connectors used in \textit{OASIS-UROS}, smaller, lighter SMB connectors were chosen for \textit{OASIS-ERIS}. While previously the constant current was adjustable using a potentiometer, the constant current sources are now fixed to a current of $\approx\SI{3}{\milli\ampere}$, which should be sufficient for most IEPE transducers. The constant current, together with the cables' capacitance, determines the maximum frequency that can be transmitted from the sensor to the acquisition system. An increase of the constant current is only required if very long cables are used~\cites{ADCN0540}{PCBLongCables}.

Further, the integrated circuits (ICs) and corresponding circuits used for the IEPE supply were updated: the required \SI{24}{\volt} supply is now generated by a \textit{Texas Instruments LM5157} boost-converter~\cite{LM5157}, and the constant current is generated by \textit{Texas Instruments LM334M} constant current sources~\cite{LM334}. For the boost converter, this change enabled a smaller overall footprint on the board without bulky electrolytic capacitors. The new constant current sources offer similar performance but are significantly cheaper. Using jumpers, it is still possible to disable the IEPE supply for each input channel. However, the required high-pass filter (passive RC filter with a \SI{-3}{\decibel} cutoff frequency of \SI{0.8}{\hertz}) cannot be disabled.

Smaller improvements include an added USB power LED, a refined on-board WiFi antenna, and the addition of an internal USB header, for instance, to integrate \textit{OASIS-ERIS} into Linux ecosystems. The ability to manually modify electrical components has been greatly improved by increasing the size of all SMD (surface mount device) components to at least 0603. Lastly, each channel now features an overload indication through an LED.

\subsection{Software features}

On the firmware side, the biggest change is the migration from the Arduino ESP32 framework to a direct C++ implementation with the ESP-IDF (Espressif IoT Development Framework)~\cite{ESPIDF}. This change enables faster, more direct hardware access, resulting in higher sample rates than the previous Arduino-based firmware. It also enables access to features not yet implemented in the Arduino ESP32 framework, most notably the option to expose the on-board micro SD card to a PC/laptop as a mass storage device (USBMSC). The key advantage of this is the increased transfer speed of sample data. Instead of relying on the serial connection with speeds of $\approx\SI[per-mode=symbol]{1}{\mega\bit\per\second}$, a USB 1.1 connection with speeds up to $\approx\SI[per-mode=symbol]{12}{\mega\bit\per\second}$ can be used.

\begin{figure}[H]
	\centering
	\scriptsize
	\def\svgwidth{\textwidth}
	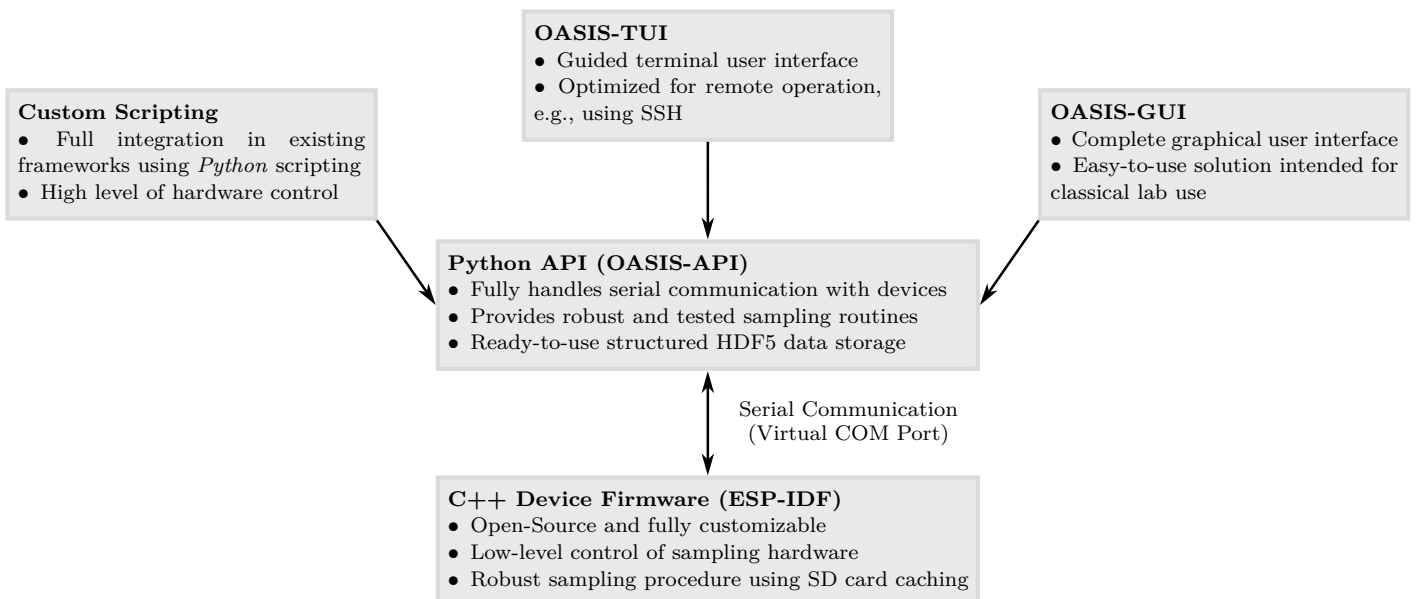
	\caption{Overview of the current \textit{OASIS} software stack.}
	\label{fig:software_stack}
\end{figure}

Further improvements were also made to the controlling software on the PC/laptop side. An overview of the current software stack is shown in \cref{fig:software_stack}.

\vspace{-0.4cm}

The communication with the \textit{OASIS} firmware over serial is now handled by a \textit{Python} API package, the \textit{OASIS-API}\footnote{\url{https://gitlab.com/oasis-acquisition/oasis-api}}. This package also provides predefined sampling routines and data storage using the HDF5\footnote{\url{https://www.hdfgroup.org/solutions/hdf5/}} format. Based on the \textit{OASIS-API}, it is possible to create custom \textit{Python} scripting routines for data acquisition without the necessity of low-level hardware control. Additionally, ready-to-use \textit{Python} packages based on the \textit{OASIS-API} are provided: for using \textit{OASIS} from a terminal, e.g., with remote access over SSH, the \textit{OASIS-TUI}\footnote{\url{https://gitlab.com/oasis-acquisition/oasis-tui}} is provided, delivering a guided terminal user interface. For more classical lab applications, a full graphical user interface, the \textit{OASIS-GUI}\footnote{\url{https://gitlab.com/oasis-acquisition/oasis-gui}} is provided. All \textit{Python} packages can be installed from the \textit{Python Package Index}\footnote{\url{https://pypi.org/}} using \textit{pip}. More details on the operation of the above-described tools can be found in \cref{sec:python}.

\subsection{Inter-device synchronization}

\begin{figure}[H]
	\centering
	\scriptsize
	\def\svgwidth{0.75\textwidth}
	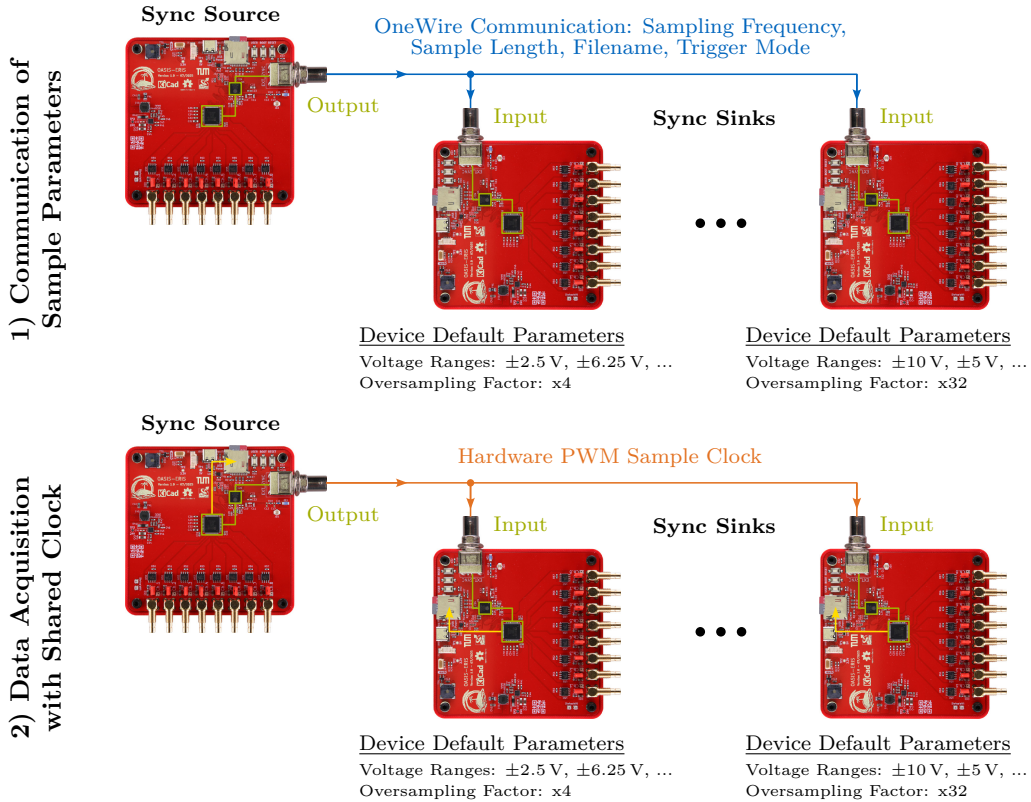
	\caption{Illustration of the \textit{OASIS} Inter-Device Synchronization feature. The \textit{Sync Source} device first communicates the sample parameters to all attached \textit{Sync Sinks}. Then, the \textit{Sync Source} transmits the hardware-generated PWM sample clock to all \textit{Sync Sink} devices. Samples are stored locally on each device and can be transferred through the USB-C port, for example, fully automatically with the provided \textit{Python} API.}
	\label{fig:sync}
\end{figure}

The flagship feature of \textit{OASIS-ERIS}, which provides a significant advancement over the previous \textit{OASIS-UROS}~\cite{OASISUROS}, is the added inter-device synchronization. Unlike other common approaches, the implemented synchronization feature does not rely on sending a synchronization pulse or on aligning the sampled data in a post-processing step; instead, it physically connects the sample clock across devices. This approach provides a truly synchronized data acquisition without any required user configuration. The working principle is explained below and illustrated in \cref{fig:sync}.

\vspace{-0.4cm}

\newpage

Arbitrarily many \textit{OASIS-ERIS} boards can be synchronized together by connecting the \textit{EXT\_SYNC} BNC connectors between all boards. This connector is attached to the \textit{CONVST} pin of the \textit{AD7606C-18} ADC and to the \textit{ESP32-S3} micro-controller. The ADC starts the analog-to-digital conversion of its inputs when a rising flank is detected on this pin. During normal sampling, the \textit{ESP32-S3} generates a sampling clock using a hardware PWM block, which triggers each ADC sample.

A synchronized sample starts with the user triggering a data acquisition on one board, which then becomes the \textit{Sync Source}, i.e., the device that controls the sample. By default, all boards are listening for \textit{OneWire} communications on the \textit{EXT\_SYNC} pin. When the sample is triggered, the \textit{Sync Source} starts transmitting the sample parameters on the \textit{EXT\_SYNC} BNC connection to the connected boards, the \textit{Sync Sinks}, using \textit{OneWire}. This communication contains the sampling frequency, the sample length, a unique filename for storage, and the trigger mode, i.e., whether a level trigger on the \textit{Sync Source} is used to start the sample. Neither voltage ranges nor oversampling factors are transmitted; instead, those parameters are taken from each board's saved default parameters.

After the sample parameters have been communicated, the \textit{Sync Source} starts the hardware PWM-generated sample clock, which is directly attached to the \textit{EXT\_SYNC} connector, transmitted through BNC cables, and also directly wired to the \textit{Sync Sinks}' ADC \textit{CONVST} pin. Therefore, the only delay between samples of the \textit{Sync Source} and the \textit{Sync Sinks} is due to the signal propagation delay between devices.

Data acquisition is complete when the \textit{Sync Source} has collected the required number of samples, as determined by the sampling frequency and duration, and shuts down the sample clock. For triggered sampling, the \textit{Sync Sources} must wait an unknown amount of time before the acquisition is triggered. During this time, all boards are already sampling data and storing it on the micro SD card. When the trigger level is reached, there is no communication of this to the \textit{Sync Sinks} since the \textit{EXT\_SYNC} connector already carries the sampling clock. Instead, when the data acquisition is finished, detected by the \textit{Sync Sinks} due to the stop of the sample clock, the already acquired data is edited such that the start aligns with the \textit{Sync Source}'s sample start. Again, this can be reconstructed from the number of samples that must be acquired, given the sampling frequency and duration.

At this point, the acquired data is stored on the SD card of each \textit{OASIS-ERIS} board. When all boards are connected to the same PC/laptop, the provided \textit{Python} software will automatically download data from the boards and combine it into a single file.
% !TeX spellcheck = en_US
\section{Design files summary}\label{sec:design_files}

A summary of the provided design files is provided in the table below. Please note that the most up-to-date versions can be found on the \textit{OASIS-Acquisition} GitLab page: \url{https://gitlab.com/oasis-acquisition}.
\vskip 0.1cm
\begin{table}[H]
	\tabulinesep=1ex
	\tiny
	\begin{tabu} to \linewidth {|X[0.65,1]|X[0.75,1]|X[0.55,1]|X|} 
		\hline
		\textbf{Design filename} & \textbf{File type} & \textbf{Open-source license} & \textbf{Location of the file} \\\hline
		%Insert design files
		oasis-eris-hardware.zip & KiCad project & CC-BY 4.0 & \url{https://doi.org/10.5281/zenodo.21426150}~\cite{OASIS_ERIS_Zenodo} \\\hline
		jlcpcb-manufacturing-files.zip & JLCPCB manufacturing files & CC-BY 4.0 & \url{https://doi.org/10.5281/zenodo.21426150}~\cite{OASIS_ERIS_Zenodo} \\\hline
		oasis-firmware.zip & C++ source code (ESP-IDF) \& pre-compiled firmware binary & MIT & \url{https://doi.org/10.5281/zenodo.21426150}~\cite{OASIS_ERIS_Zenodo} \\\hline
		oasis-factory-flasher.zip & Python source code \& pre-compiled binaries for Windows, MacOS, and Linux & MIT & \url{https://doi.org/10.5281/zenodo.21426150}~\cite{OASIS_ERIS_Zenodo} \\\hline
		oasis-gui.zip & Python package & MIT & \url{https://doi.org/10.5281/zenodo.21426150}~\cite{OASIS_ERIS_Zenodo} \\\hline
		oasis-tui.zip & Python package & MIT & \url{https://doi.org/10.5281/zenodo.21426150}~\cite{OASIS_ERIS_Zenodo} \\\hline
		oasis-api.zip & Python package & MIT & \url{https://doi.org/10.5281/zenodo.21426150}~\cite{OASIS_ERIS_Zenodo} \\\hline
		eris\_bottom\_case.stl & STL file for 3D printing & CC-BY 4.0 & \url{https://doi.org/10.5281/zenodo.21426150}~\cite{OASIS_ERIS_Zenodo} \\\hline
	\end{tabu}
	\caption{Summary of design files provided in the Zenodo repository.}
\end{table}\vspace{-0.4cm}

\noindent\textbf{oasis-eris-hardware.zip}\hspace{0.2cm}This archive contains the full \textit{KiCad}\footnote{\url{https://www.kicad.org/}} project of the \textit{OASIS-ERIS} board as described here. Included are the \textit{KiCad} schematics, the \textit{KiCad} PCB, and all required symbols and footprints.

\noindent\textbf{jlcpcb-manufacturing-files.zip}\hspace{0.2cm}Included here are the production files used by \textit{JLCPCB} to fabricate the printed circuit board. Using those files, the same board as used for validation can be ordered.

\noindent\textbf{oasis-firmware.zip}\hspace{0.2cm}In this archive, the source code for the board firmware as well as a pre-compiled binary can be found. Instructions for uploading the firmware to the board can be found in \cref{sec:bootloader_firmware}.

\noindent\textbf{oasis-factory-flasher.zip}\hspace{0.2cm}This archive contains the Python source code as well as pre-compiled binaries for the \textit{OASIS-FactoryFlasher} (see also \cref{sec:bootloader_firmware}). This program is used for the initial firmware setup and \textit{ESP32} configuration of the \textit{OASIS} board.

\noindent\textbf{oasis-gui.zip}\hspace{0.2cm}For archival purposes, the source code and the built \textit{Python} package for the \textit{OASIS-GUI} (see also \cref{sec:gui}) is included in the design files. However, it is recommended that the current version from the \textit{Python Package Index} is installed using \texttt{pip install OASIS-GUI}\footnote{\url{https://pypi.org/project/oasis-gui/}}. 

\noindent\textbf{oasis-tui.zip}\hspace{0.2cm}For archival purposes, the source code and the built \textit{Python} package for the \textit{OASIS-TUI} (see also \cref{sec:tui}) is included in the design files. However, it is recommended that the current version from the \textit{Python Package Index} is installed using \texttt{pip install OASIS-TUI}\footnote{\url{https://pypi.org/project/oasis-tui/}}. 

\noindent\textbf{oasis-api.zip}\hspace{0.2cm}For archival purposes, the source code and the built \textit{Python} package for the \textit{OASIS-API} (see also \cref{sec:api}) is included in the design files. However, it is recommended that the current version from the \textit{Python Package Index} is installed using \texttt{pip install OASIS-API}\footnote{\url{https://pypi.org/project/oasis-api/}}. 

\noindent\textbf{eris\_bottom\_case.stl}\hspace{0.2cm}Lastly, for protection against mechanical damage, shielding the bottom of the finished board is recommended. A 3D printing file for a simple case is included with this file.
% !TeX spellcheck = en_US
\section{Bill of materials summary}

The bill of materials for \textit{OASIS-ERIS} is split into two parts: \cref{tab:ERIS_SMD} lists the bare printed circuit board (PCB) and all surface mount devices (SMD) that the PCB manufacturer can directly assemble. \Cref{tab:ERIS_manual} lists the remaining components that should be assembled by hand, namely the BNC and SMB connectors, as well as some jumpers.

% JLCPCB Parts
\begin{table}[H]
	\tabulinesep=0.4ex
	\noindent{\tiny
	\begin{tabu} to \linewidth {|X[0.45,1]|X[1.5,1]|X[0.2,1]|X[0.3,1]|X[0.3,1]|X[0.65,1]|X[0.4,1]|}
		\hline
		\textbf{Designator} & \textbf{Component} & \textbf{Number} & \textbf{Cost per unit - currency} & \textbf{Total cost - currency} & \textbf{Source of materials} & \textbf{Material type} \\\hline
		PCB & Bare printed circuit board & 1 & $\approx$ $20$\;\euro & $\approx$ $20$\;\euro & \href{https://jlcpcb.com/}{JLCPCB} & Other \\\hline
		ADC0 &  Analog-Digital Converter AD7606C-18BSTZ & 1 & $42.51$\;\euro & $42.51$\;\euro & \href{https://jlcpcb.com/partdetail/C2924606}{C2924606 (JLCPCB)} & Semiconductor \\\hline
		AE1 & Antenna RFANT3216120A5T & 1 & $0.05$\;\euro & $0.05$\;\euro & \href{https://jlcpcb.com/partdetail/C127629}{C127629 (JLCPCB)} & Semiconductor \\\hline
		C10, C11 & Capacitor 1pF 16V X5R & 2 & $0.0042$\;\euro & $0.0084$\;\euro & \href{https://jlcpcb.com/partdetail/C115062}{C115062 (JLCPCB)} & Semiconductor \\\hline
		C20, C210, C34, C36, C44 & Capacitor 1uF 16V X5R & 5 & $0.0047$\;\euro & $0.0235$\;\euro & \href{https://jlcpcb.com/partdetail/C15849}{C15849 (JLCPCB)} & Semiconductor \\\hline
		C21, C23, C24, C25, C27, C30, C31, C33, C35, C43 & Capacitor 100nF 16V X5R & 10 & $0.0022$\;\euro & $0.0220$\;\euro & \href{https://jlcpcb.com/partdetail/C14663}{C14663 (JLCPCB)} & Semiconductor \\\hline
		C22, C32, C41 & Capacitor 10uF 16V X5R & 3 & $0.0555$\;\euro & $0.1665$\;\euro & \href{https://jlcpcb.com/partdetail/C440198}{C440198 (JLCPCB)} & Semiconductor \\\hline
		C26, C90, C91 & Capacitor 4.7uF 16V X5R & 3 & $0.0089$\;\euro & $0.0267$\;\euro & \href{https://jlcpcb.com/partdetail/C19666}{C19666 (JLCPCB)} & Semiconductor \\\hline
		C28, C29 & Capacitor 10pF 16V X5R & 2 & $0.0039$\;\euro & $0.0078$\;\euro & \href{https://jlcpcb.com/partdetail/C1634}{C1634 (JLCPCB)} & Semiconductor \\\hline
		C42 & Capacitor 12nF 16V X5R & 1 & $0.0029$\;\euro & $0.0029$\;\euro & \href{https://jlcpcb.com/partdetail/C107074}{C107074 (JLCPCB)} & Semiconductor \\\hline
		C45 & Capacitor 5.6nF 16V X5R & 1 & $0.0124$\;\euro & $0.0124$\;\euro & \href{https://jlcpcb.com/partdetail/C440183}{C440183 (JLCPCB)} & Semiconductor \\\hline
		C46 & Capacitor 360pF 16V X5R & 1 & $0.0041$\;\euro & $0.0041$\;\euro & \href{https://jlcpcb.com/partdetail/C513587}{C513587 (JLCPCB)} & Semiconductor \\\hline
		C47, C48 & Capacitor 4.7uF 50V X5R & 2 & $0.0320$\;\euro & $0.0640$\;\euro & \href{https://jlcpcb.com/partdetail/C29823}{C29823 (JLCPCB)} & Semiconductor \\\hline
		C49 & Capacitor 100nF 50V X5R & 1 & $0.0022$\;\euro & $0.0022$\;\euro & \href{https://jlcpcb.com/partdetail/C14663}{C14663 (JLCPCB)} & Semiconductor \\\hline
		C60-C67 & Capacitor 10uF 50V X5R & 8 & $0.0555$\;\euro & $0.4440$\;\euro & \href{https://jlcpcb.com/partdetail/C440198}{C440198 (JLCPCB)} & Semiconductor \\\hline
		D1 & White LED & 1 & $0.0099$\;\euro & $0.0099$\;\euro & \href{https://jlcpcb.com/partdetail/C2290}{C2290 (JLCPCB)} & Semiconductor \\\hline
		D41 & DSK14 Schottky Diode & 1 & $0.0141$\;\euro & $0.0141$\;\euro & \href{https://jlcpcb.com/partdetail/C908228}{C908228 (JLCPCB)} & Semiconductor \\\hline
		F1 & Fuse JK-nSMD050-30 & 1 & $0.0274$\;\euro & $0.0274$\;\euro & \href{https://jlcpcb.com/partdetail/C720075}{C720075 (JLCPCB)} & Semiconductor \\\hline
		J1 & MOLEX Connector 533980471 & 1 & $0.1557$\;\euro & $0.1557$\;\euro & \href{https://jlcpcb.com/partdetail/C17617036}{C17617036 (JLCPCB)} & Semiconductor \\\hline
		J4 & JST Connector BM04B-SRSS-TB(LF)(SN) & 1 & $0.1728$\;\euro & $0.1728$\;\euro & \href{https://jlcpcb.com/partdetail/C160390}{C160390 (JLCPCB)} & Semiconductor \\\hline
		J5 & microSD Connector GT-TF003-H0185-02 & 1 & $0.1596$\;\euro & $0.1596$\;\euro & \href{https://jlcpcb.com/partdetail/C5155564}{C5155564 (JLCPCB)} & Semiconductor \\\hline
		J10 & Antenna Connector U.FL-R-SMT-1(80) & 1 & $0.1137$\;\euro & $0.1137$\;\euro & \href{https://jlcpcb.com/partdetail/C88374}{C88374 (JLCPCB)} & Semiconductor \\\hline
		J80 & USB-C Connector GT-USB-7010ASV & 1 & $0.0712$\;\euro & $0.0712$\;\euro & \href{https://jlcpcb.com/partdetail/C2988369}{C2988369 (JLCPCB)} & Semiconductor \\\hline
		L10 & Inductor 2nH & 1 & $0.0093$\;\euro & $0.0093$\;\euro & \href{https://jlcpcb.com/partdetail/C395081}{C395081 (JLCPCB)} & Semiconductor \\\hline
		L11 & Inductor 6.8nH & 1 & $0.0313$\;\euro & $0.0313$\;\euro & \href{https://jlcpcb.com/partdetail/C86136}{C86136 (JLCPCB)} & Semiconductor \\\hline
		L41 & Inductor 1.8uH & 1 & $0.4645$\;\euro & $0.4645$\;\euro & \href{https://jlcpcb.com/partdetail/C2045226}{C2045226 (JLCPCB)} & Semiconductor \\\hline
		LED1-LED10 & RGB LED WS2812C-2020-V1 & 10 & $0.0718$\;\euro & $0.7180$\;\euro & \href{https://jlcpcb.com/partdetail/C2976072}{C2976072 (JLCPCB)} & Semiconductor \\\hline
		LS1 & Buzzer KSSG74B16 & 1 & $0.7068$\;\euro & $0.7068$\;\euro & \href{https://jlcpcb.com/partdetail/C7437070}{C7437070 (JLCPCB)} & Semiconductor \\\hline
		R1-R7, R22-R24 & Resistor 10k & 10 & $0.1001$\;\euro & $1.0010$\;\euro & \href{https://jlcpcb.com/partdetail/C2074224}{C2074224 (JLCPCB)} & Semiconductor \\\hline
		R8 & Resistor 680R & 1 & $0.0011$\;\euro & $0.0011$\;\euro & \href{https://jlcpcb.com/partdetail/C23228}{C23228 (JLCPCB)} & Semiconductor \\\hline
		R10 & Resistor 0R & 1 & $0.0010$\;\euro & $0.0010$\;\euro & \href{https://jlcpcb.com/partdetail/C21189}{C21189 (JLCPCB)} & Semiconductor \\\hline
		R20, R21 & Resistor 33R & 2 & $0.0014$\;\euro & $0.0028$\;\euro & \href{https://jlcpcb.com/partdetail/C23140}{C23140 (JLCPCB)} & Semiconductor \\\hline
		R41 & Resistor 18.7k & 1 & $0.0013$\;\euro & $0.0013$\;\euro & \href{https://jlcpcb.com/partdetail/C155690}{C155690 (JLCPCB)} & Semiconductor \\\hline
		R42 & Resistor 100k & 1 & $0.0009$\;\euro & $0.0009$\;\euro & \href{https://jlcpcb.com/partdetail/C25803}{C25803 (JLCPCB)} & Semiconductor \\\hline
		R43 & Resistor 4.99k & 1 & $0.0010$\;\euro & $0.0010$\;\euro & \href{https://jlcpcb.com/partdetail/C23046}{C23046 (JLCPCB)} & Semiconductor \\\hline
		R44 & Resistor 57.6k & 1 & $0.0014$\;\euro & $0.0014$\;\euro & \href{https://jlcpcb.com/partdetail/C245983}{C245983 (JLCPCB)} & Semiconductor \\\hline
		R45 & Resistor 2.49k & 1 & $0.0236$\;\euro & $0.0236$\;\euro & \href{https://jlcpcb.com/partdetail/C705842}{C705842 (JLCPCB)} & Semiconductor \\\hline
		R46 & Resistor 5.1R & 1 & $0.0018$\;\euro & $0.0018$\;\euro & \href{https://jlcpcb.com/partdetail/C25197}{C25197 (JLCPCB)} & Semiconductor \\\hline
		R50-R57 & Resistor 22R & 8 & $0.0009$\;\euro & $0.0072$\;\euro & \href{https://jlcpcb.com/partdetail/C23345}{C23345 (JLCPCB)} & Semiconductor \\\hline
		R60-R67 & Resistor 20k & 8 & $0.0010$\;\euro & $0.0080$\;\euro & \href{https://jlcpcb.com/partdetail/C4184}{C4184 (JLCPCB)} & Semiconductor \\\hline
		R70 & Resistor 1k & 1 & $0.0010$\;\euro & $0.0010$\;\euro & \href{https://jlcpcb.com/partdetail/C21190}{C21190 (JLCPCB)} & Semiconductor \\\hline
		R80, R81 & Resistor 5.1k & 2 & $0.0012$\;\euro & $0.0024$\;\euro & \href{https://jlcpcb.com/partdetail/C23186}{C23186 (JLCPCB)} & Semiconductor \\\hline
		R90 & Resistor 10k & 1 & $0.0011$\;\euro & $0.0011$\;\euro & \href{https://jlcpcb.com/partdetail/C25804}{C25804 (JLCPCB)} & Semiconductor \\\hline
		SW1-SW3 & Switch TS-1101-C-W & 3 & $0.0361$\;\euro & $0.1083$\;\euro & \href{https://jlcpcb.com/partdetail/C318938}{C318938 (JLCPCB)} & Semiconductor \\\hline
		T70 & Transistor DDTC114YCA-7-F & 1 & $0.0421$\;\euro & $0.0421$\;\euro & \href{https://jlcpcb.com/partdetail/C57530}{C57530 (JLCPCB)} & Semiconductor \\\hline
		U41 & Boost Converter LM5157QRTERQ1 & 1 & $2.3867$\;\euro & $2.3867$\;\euro & \href{https://jlcpcb.com/partdetail/C3188429}{C3188429 (JLCPCB)} & Semiconductor \\\hline
		U50-U57 & Constant Current Source LM334M & 8 & $0.5328$\;\euro & $4.2624$\;\euro & \href{https://jlcpcb.com/partdetail/C7945}{C7945 (JLCPCB)} & Semiconductor \\\hline
		U80 & Surge Protection USBLC6-2P6 & 1 & $0.1557$\;\euro & $0.1557$\;\euro & \href{https://jlcpcb.com/partdetail/C15999}{C15999 (JLCPCB)} & Semiconductor \\\hline
		U90 & 3.3V Regulator TLV76733DRVR & 1 & $0.1687$\;\euro & $0.1687$\;\euro & \href{https://jlcpcb.com/partdetail/C2848334}{C2848334 (JLCPCB)} & Semiconductor \\\hline
		Y20 & 40MHz Crystal Oscillator SX2B40.000F1210F30 & 1 & $0.0893$\;\euro & $0.0893$\;\euro & \href{https://jlcpcb.com/partdetail/C2901733}{C2901733 (JLCPCB)} & Semiconductor \\\hline
	\end{tabu}}
	\caption{Bill of materials for the \textit{OASIS-ERIS} board, not including connectors to be soldered manually. Prices as of 20.02.2026.}
	\label{tab:ERIS_SMD}
\end{table}\vspace{-0.4cm}

% Manual assembly
\begin{table}[H]
	\tabulinesep=0.4ex
	\noindent{\tiny
		\begin{tabu} to \linewidth {|X[0.45,1]|X[1.5,1]|X[0.2,1]|X[0.3,1]|X[0.3,1]|X[0.65,1]|X[0.4,1]|}
			\hline
			\textbf{Designator} & \textbf{Component} & \textbf{Number} & \textbf{Cost per unit - currency} & \textbf{Total cost - currency} & \textbf{Source of materials} & \textbf{Material type} \\\hline
			EXT\_SYNC & BNC Connector 1-1337481-0  & 1 & $2.18$\;\euro & $2.18$\;\euro & \href{https://www.mouser.com/ProductDetail/712-CONBNC002}{712-CONBNC002 (Mouser)} & Other \\\hline
			X1-X8 & SMB Connector TE Connectivity 1-1337481-0 & 8 & $6.47$\;\euro & $51.76$\;\euro & \href{https://www.mouser.com/ProductDetail/571-1-1337481-0}{571-1-1337481-0 (Mouser)} & Other\\\hline
			JP1-JP8 & Jumper \& Pinheaders & 8 & $0.1940$\;\euro & $1.552$\;\euro & \href{https://www.mouser.com/ProductDetail/571-28815452}{571-28815452 (Mouser)} & Other \\\hline
	\end{tabu}}
	\caption{Components required for final manual assembly of the \textit{OASIS-ERIS} board. Prices as of 20.02.2026.}
	\label{tab:ERIS_manual}
\end{table}\vspace{-0.4cm}
% !TeX spellcheck = en_US
\section{Build instructions}

In this section, the hardware assembly of \textit{OASIS-ERIS} is explained in detail. Then, the required firmware setup and device configuration is covered. The setup of the PC/laptop Python software are covered in \cref{sec:python}.

\subsection{Hardware assembly}

\begin{figure}[H]
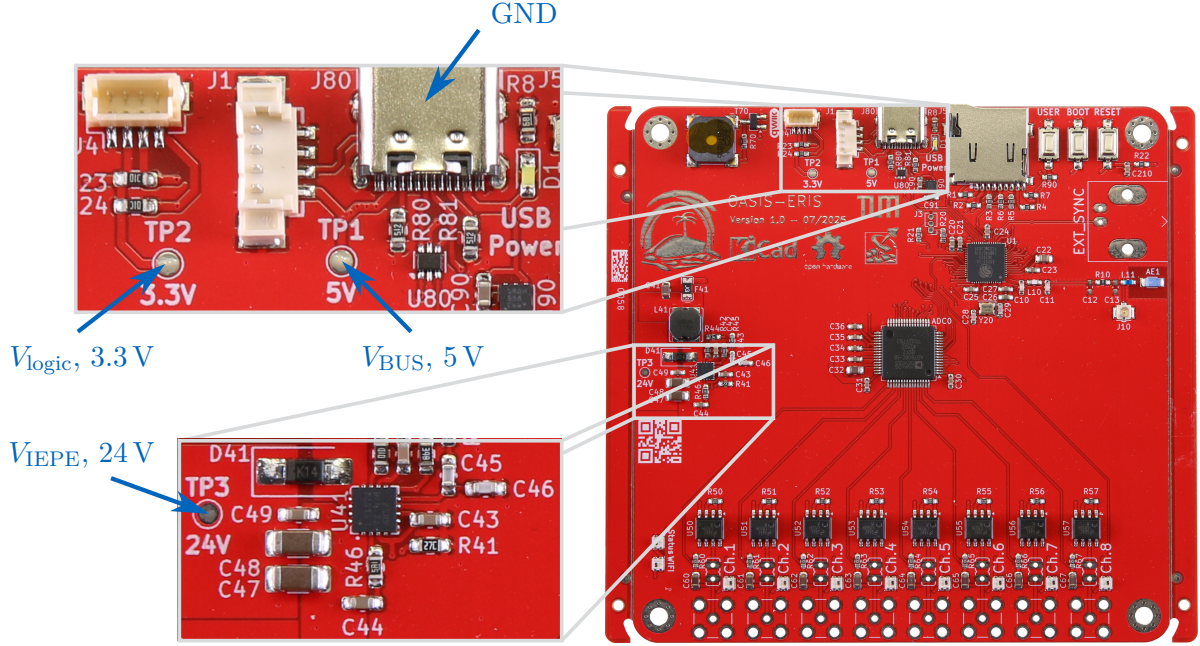

	\centering
	\def\svgwidth{0.8\textwidth}
	%% Creator: Inkscape 1.4.4 (dcaf3e7d9e, 2026-05-05), www.inkscape.org
%% Accompanies image file 'jlcpcb_raw.pdf' (pdf, eps, ps)
\begingroup%
  \makeatletter%
  \providecommand\color[2][]{%
    \errmessage{(Inkscape) Color is used for the text in Inkscape, but the package 'color.sty' is not loaded}%
    \renewcommand\color[2][]{}%
  }%
  \providecommand\transparent[1]{%
    \errmessage{(Inkscape) Transparency is used (non-zero) for the text in Inkscape, but the package 'transparent.sty' is not loaded}%
    \renewcommand\transparent[1]{}%
  }%
  \providecommand\rotatebox[2]{#2}%
  \newcommand*\fsize{\dimexpr\f@size pt\relax}%
  \newcommand*\lineheight[1]{\fontsize{\fsize}{#1\fsize}\selectfont}%
  \ifx\svgwidth\undefined%
    \setlength{\unitlength}{5918.34361159bp}%
    \ifx\svgscale\undefined%
      \relax%
    \else%
      \setlength{\unitlength}{\unitlength * \real{\svgscale}}%
    \fi%
  \else%
    \setlength{\unitlength}{\svgwidth}%
  \fi%
  \global\let\svgwidth\undefined%
  \global\let\svgscale\undefined%
  \makeatother%
  \begin{picture}(1,0.5690607)%
    \lineheight{1}%
    \setlength\tabcolsep{0pt}%
    \put(0,0){\includegraphics[width=\unitlength,page=1]{jlcpcb_raw.pdf}}%
    \put(0.31053345,0.24611089){\color[rgb]{0,0.39607843,0.74117647}\makebox(0,0)[t]{\lineheight{1.25}\smash{\begin{tabular}[t]{c}$V_\text{BUS}$, \SI{5}{\volt}\end{tabular}}}}%
    \put(0.00723716,0.24611097){\color[rgb]{0,0.39607843,0.74117647}\makebox(0,0)[t]{\lineheight{1.25}\smash{\begin{tabular}[t]{c}$V_\text{logic}$, \SI{3.3}{\volt}\end{tabular}}}}%
    \put(0.37085059,0.55100244){\color[rgb]{0,0.39607843,0.74117647}\makebox(0,0)[lt]{\lineheight{1.25}\smash{\begin{tabular}[t]{l}GND\end{tabular}}}}%
    \put(0,0){\includegraphics[width=\unitlength,page=2]{jlcpcb_raw.pdf}}%
    \put(0.07207256,0.16091006){\color[rgb]{0,0.39607843,0.74117647}\makebox(0,0)[rt]{\lineheight{1.25}\smash{\begin{tabular}[t]{r}$V_\text{IEPE}$, \SI{24}{\volt}\end{tabular}}}}%
    \put(0,0){\includegraphics[width=\unitlength,page=3]{jlcpcb_raw.pdf}}%
  \end{picture}%
\endgroup%

	\caption{Partially assembled \textit{OASIS-ERIS} PCB using the provided JLCPCB manufacturing file (right) and recommended voltage supply rail testpoints \textbf{TP1} ($V_\text{BUS}$, \SI{5}{\volt}), \textbf{TP2} ($V_\text{logic}$, \SI{3.3}{\volt}), and \textbf{TP3} ($V_\text{IEPE}$, \SI{24}{\volt}).}
	\label{fig:JLCPCB_raw}
\end{figure}

Using the provided \textit{KiCad} project, or the \textit{JLCPCB} manufacturing files, the \textit{OASIS-ERIS} printed circuit board (PCB) can be ordered. It is recommended to let the PCB manufacturer assemble all SMD components of the board; then, only the BNC and SMB connectors, as well as the jumper headers, must be assembled manually. The partially assembled PCB, using the provided files, is shown in \cref{fig:JLCPCB_raw}.

\vspace{-0.4cm}

Before connecting the board to a PC/laptop, the board should be checked for short circuits. For this, the provided test points can be used to check the resistance versus ground: test point \textbf{TP1} ($V_\text{BUS}$, \SI{5}{\volt}) is the voltage rail of the connected USB, \textbf{TP2} ($V_\text{logic}$, \SI{3.3}{\volt}) is the voltage supply for the \textit{ESP32-S3} micro-controller, and \textbf{TP3} ($V_\text{IEPE}$, \SI{24}{\volt}) is the output of the voltage boost-converter required for the IEPE supply. If none of the supply rails are shorted, the board can be connected to a USB-C power supply. Then, the voltages at the test point should be checked to see if they match their nominal voltages. Note that the USB specification allows for $V_\text{BUS}$ to be between 4.75 and \SI{5.25}{\volt}~\cite{USBImplementersForum2000}, and that there might also be some slight deviations for the other voltages.

\subsubsection{Soldering of through-hole components}

\begin{figure}[H]
	\includegraphics[width=0.5\textwidth]{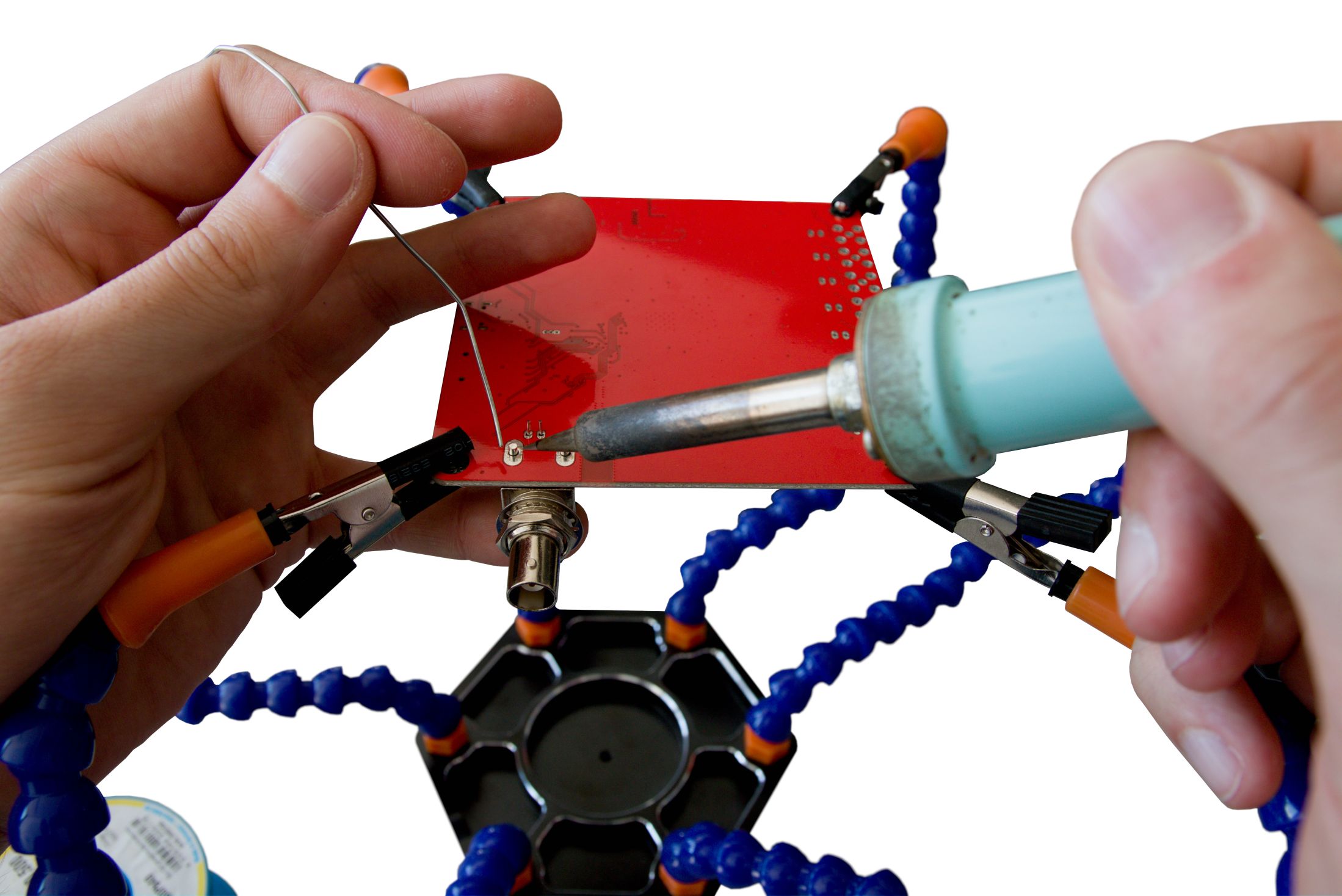}
	\includegraphics[width=0.5\textwidth]{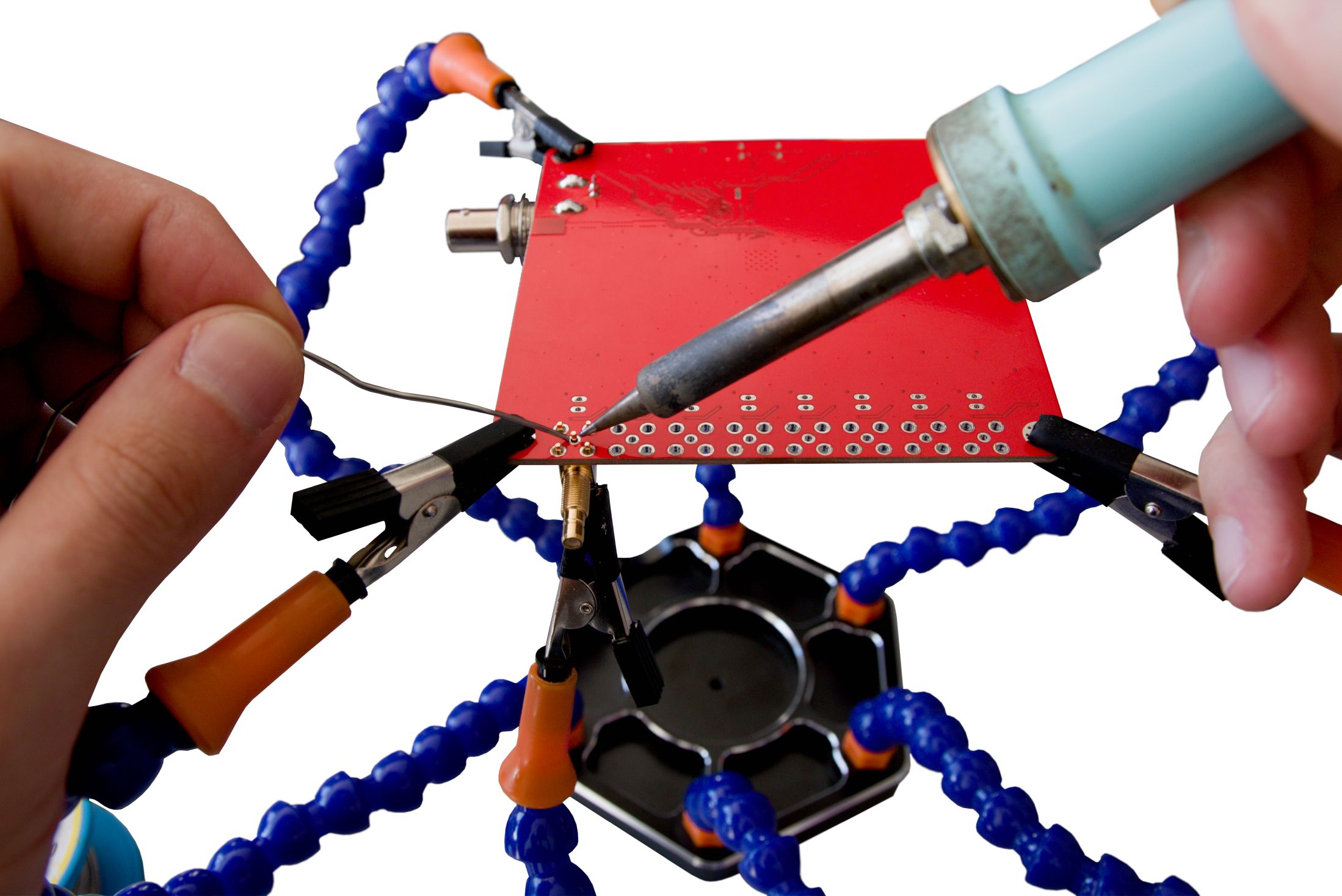}
	\caption{Soldering of through-hole connectors; on the left: BNC connector, on the right: SMB connector}
	\label{fig:assembly1}
\end{figure}

Manual assembly is required for the SMB and BNC connectors, as well as the jumper headers, used to activate or deactivate the IEPE supply. The connectors are best soldered with some tool that suspends the board in the air; see also \cref{fig:assembly1}. Due to the high heat capacity of the BNC connector, it can be held in place by hand while tacking one of the mounting pins; see the left side of \cref{fig:assembly1}. After this, the rest of the pins can be soldered without needing to hold the connector in place.

\vspace{-0.4cm}

For the eight SMB connectors, it is not possible to fix them by hand for the tacking due to them quickly becoming too hot to touch. Here, it is recommended to hold them loosely in place with a tool (see, for example, the right of \cref{fig:assembly1}) and tack one pin of the connector. After this, the connection can be reheated while pushing the connector to properly align it in the correct orientation. Then, the remaining pins can be soldered.

\begin{figure}[H]
	\includegraphics[width=0.3\textwidth]{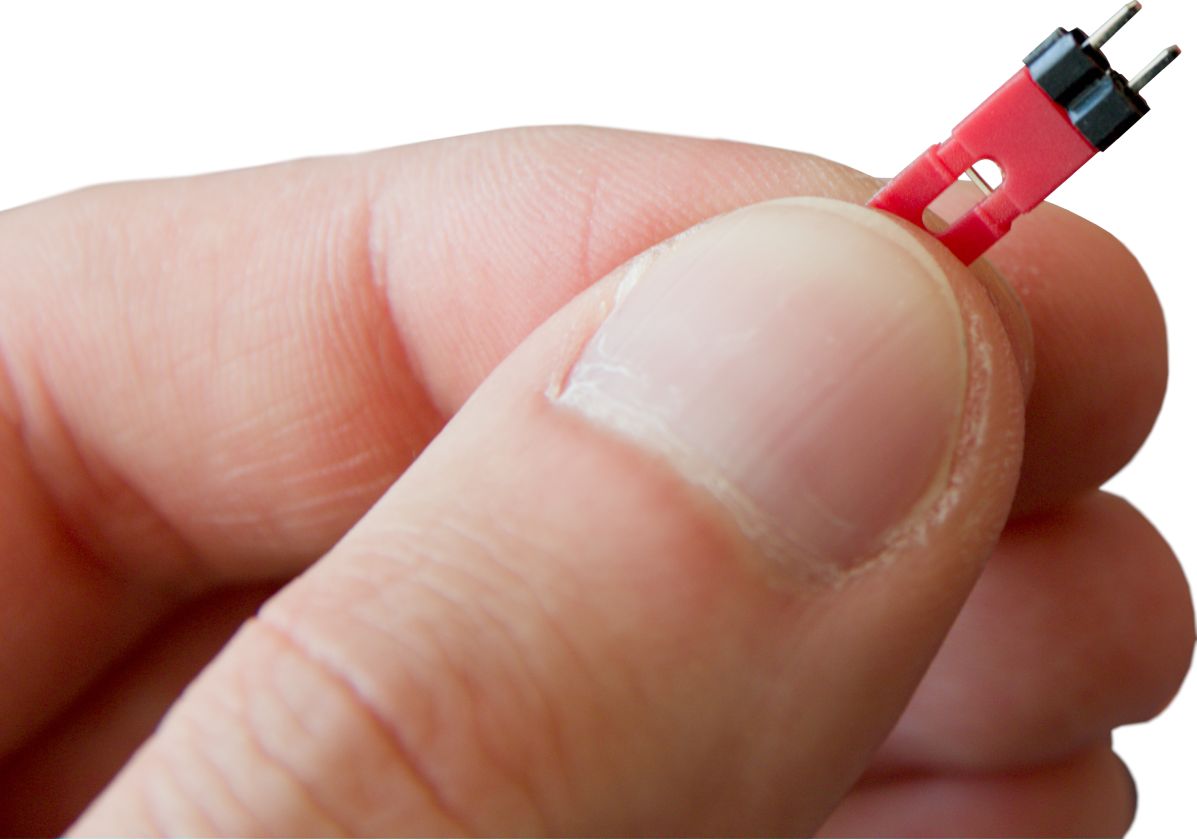}
	\includegraphics[width=0.7\textwidth]{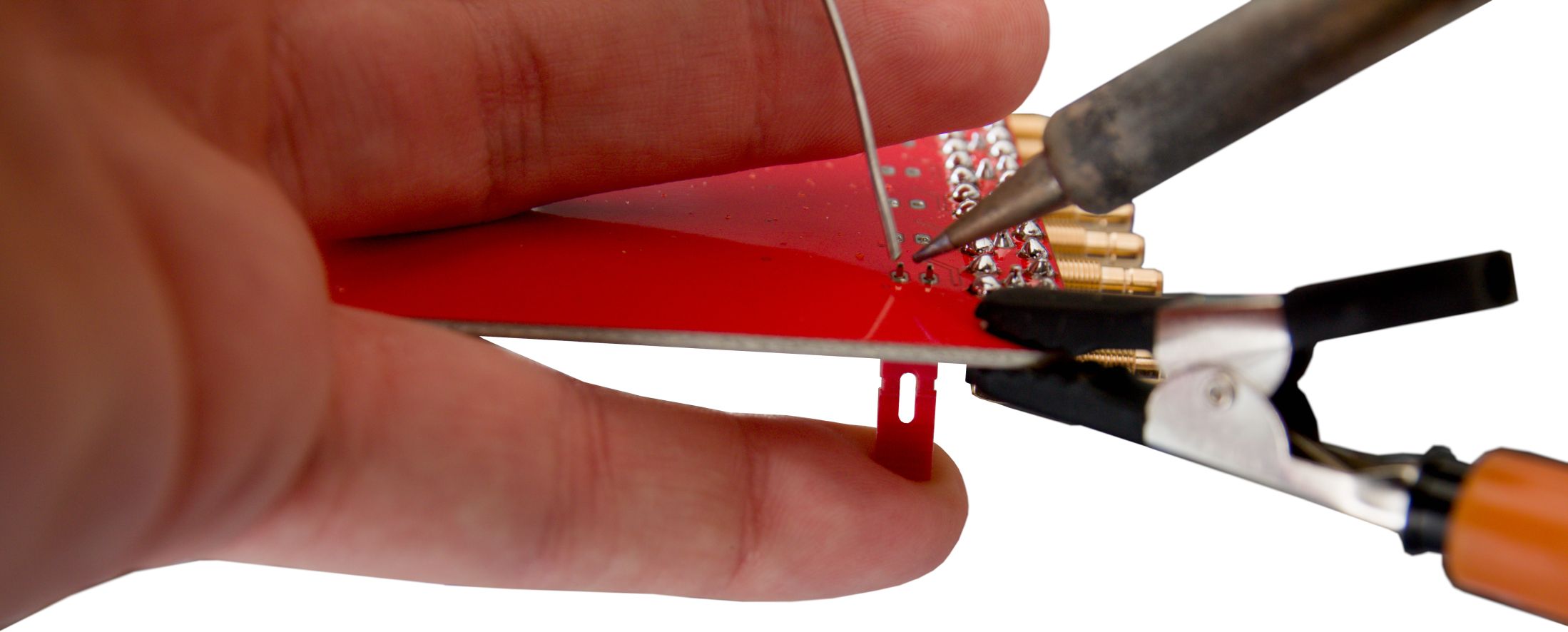}
	\caption{Soldering of pin headers by inserting them into a jumper as a handling aid.}
	\label{fig:assembly2}
\end{figure}

Lastly, the eight pin headers for the jumper must be soldered. This is best done by inserting the jumper header into a jumper (see the left of \cref{fig:assembly2}), and then holding the jumper in place by hand for tacking (see the right of \cref{fig:assembly2}). Again, after the initial tacking, the solder connection can be reheated while pushing on the jumper to properly orient it. These steps conclude the assembly of the electrical components.

\vspace{-0.4cm}

\subsubsection{Housing and mounting}

\begin{figure}[H]
	\scriptsize
	\def\svgwidth{\textwidth}
	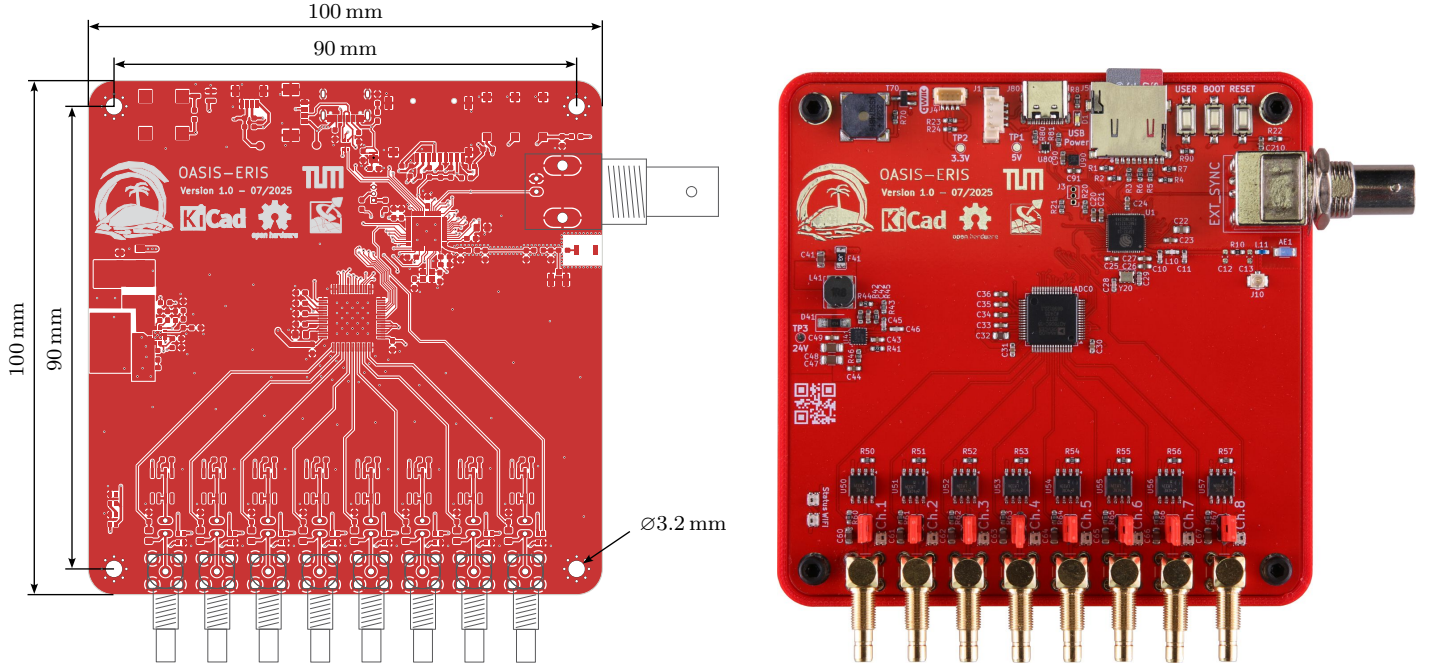
	\caption{Overview of the \textit{OASIS-ERIS} mounting dimensions (left) and assembled \textit{OASIS-ERIS} with bottom case (right).}
	\label{fig:housing_mounting}
\end{figure}

Lastly, housing and mounting options should be considered for \textit{OASIS-ERIS}. For fixed installation, \textit{OASIS-ERIS} provides four mounting holes in a $90\times\SI{90}{\milli\meter}$ grid with a diameter of \SI{3.2}{\milli\meter} for M3 screws; see also \cref{fig:housing_mounting} on the left. The conductive plate around the mounting holes is connected to the device ground.

If \textit{OASIS-ERIS} is used without a permanent installation, it is recommended to cover the bottom in order to isolate the exposed pins of the BNC and SMB connectors. For this, a cover plate can be 3D-printed using the provided file \texttt{eris\_bottom\_case.stl}; see also the design files summary in \cref{sec:design_files} or the \textit{OASIS-ERIS} repository~\cite{OASIS_ERIS_Zenodo}. The board can be fixed to the bottom case using M3 screws; see also \cref{fig:housing_mounting} on the right.

\vspace{-0.4cm}

\subsection{Firmware and device setup}

To make \textit{OASIS-ERIS} operational, some further steps are required. First, a micro SD card must be formatted using \textbf{exFAT} and inserted into the board. Next, the bootloader must be burned onto the \textit{ESP32-S3} and the \textit{OASIS-Firmware} must be uploaded. Lastly, some device metadata can be set to help identify the board.

\subsubsection{Preparing the micro SD card}

\textit{OASIS-ERIS} requires a micro SD card already formatted as \textbf{exFAT} to function correctly. Recommended are micro SD cards with at least 2 GB of storage and a speed Class of 10 or better. Step-by-step instructions:\vspace{-0.4cm}
\begin{enumerate}
	\item Insert the micro SD card into your computer or laptop.\vspace{-0.2cm}
	\item Format the card as \textbf{exFAT}:\vspace{-0.1cm}
	\begin{itemize}
		\item Windows: Open 'This PC' in File Explorer, right click the SD card, select 'Format...', select \textbf{exFAT}, and click 'Start'.
		\item macOS: Right-click the SD card on the desktop, select 'Erase Disk...', select \textbf{ExFAT}, and click 'Erase'.
		\item Linux: Open \textit{KDE Partition Manager} or \textit{GParted}, select the SD card, unmount the card if it was mounted, delete any existing partition, create a new partition with \textbf{exFAT} at the unallocated space, press 'OK', and then press 'Apply'.
	\end{itemize}
	\item Insert the formatted SD card into the \textit{OASIS} board.
\end{enumerate}\vspace{-0.4cm}

\subsubsection{Firmware compilation, burning the bootloader, and flashing the firmware}\label{sec:bootloader_firmware}

The \textit{OASIS-ERIS} firmware is based on \textit{ESP-IDF}\footnote{\url{https://docs.espressif.com/projects/esp-idf/en/stable/esp32/index.html}} version 6.0.0. To compile the firmware locally, the \textit{ESP-IDF} toolchain must be installed and configured. Since this is somewhat complicated, it is recommended instead to use the pre-compiled binary available on the \textit{GitLab} release page: \url{https://gitlab.com/oasis-acquisition/oasis-firmware/-/releases}, or, alternatively, the archived build in the \textit{OASIS-ERIS} repository~\cite{OASIS_ERIS_Zenodo}.

\begin{figure}[H]
	\centering
	\begin{subfigure}{0.31\textwidth}
		\centering
		\textbf{Windows}\vspace{0.2cm}
		\includegraphics[height=4.2cm]{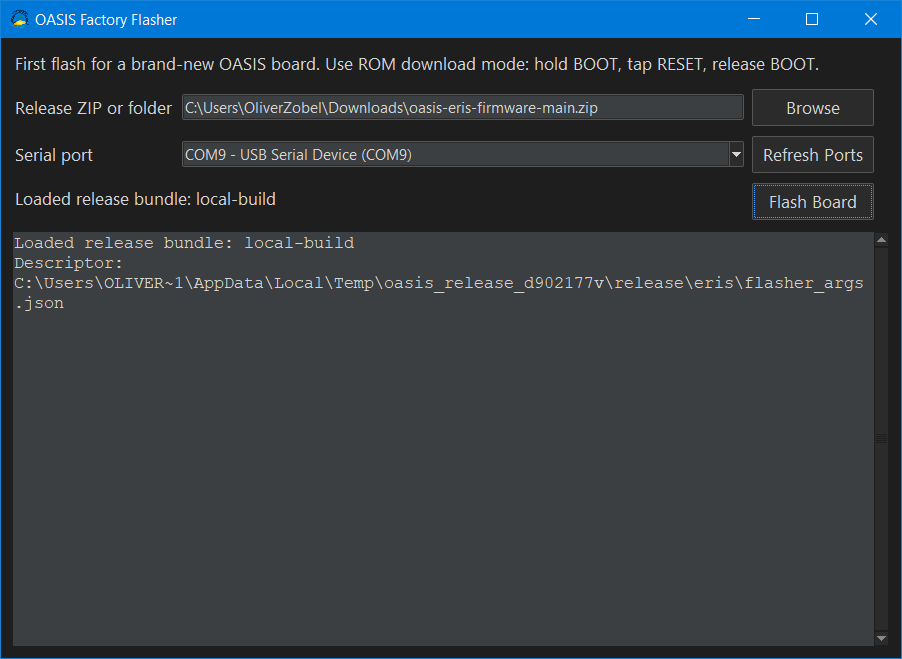}
	\end{subfigure}
	\begin{subfigure}{0.32\textwidth}
		\centering
		\textbf{MacOS}\vspace{0.2cm}
		\includegraphics[height=4.2cm]{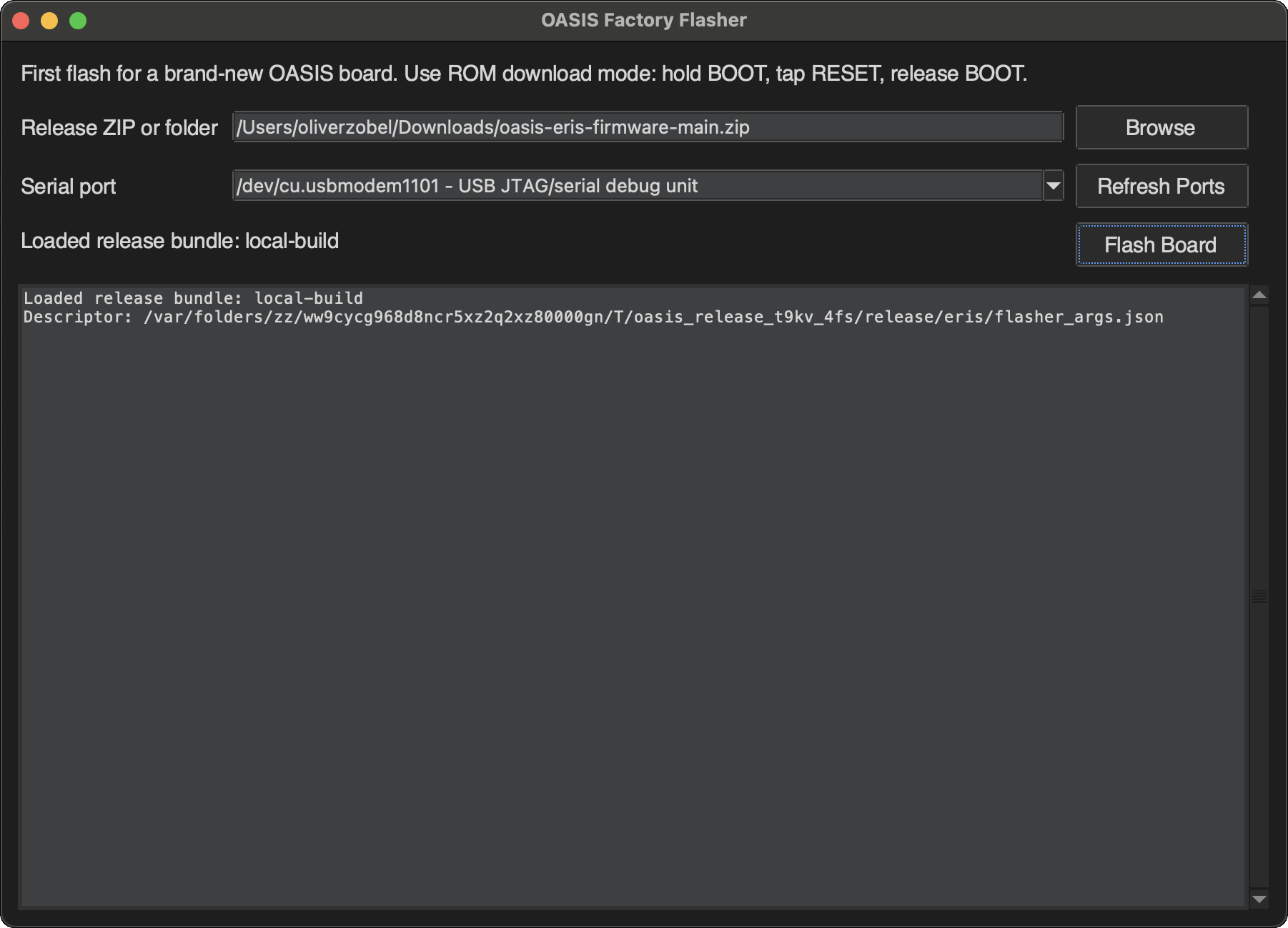}
	\end{subfigure}
	\begin{subfigure}{0.33\textwidth}
		\centering
		\textbf{Linux}\vspace{0.2cm}
		\includegraphics[height=4.2cm]{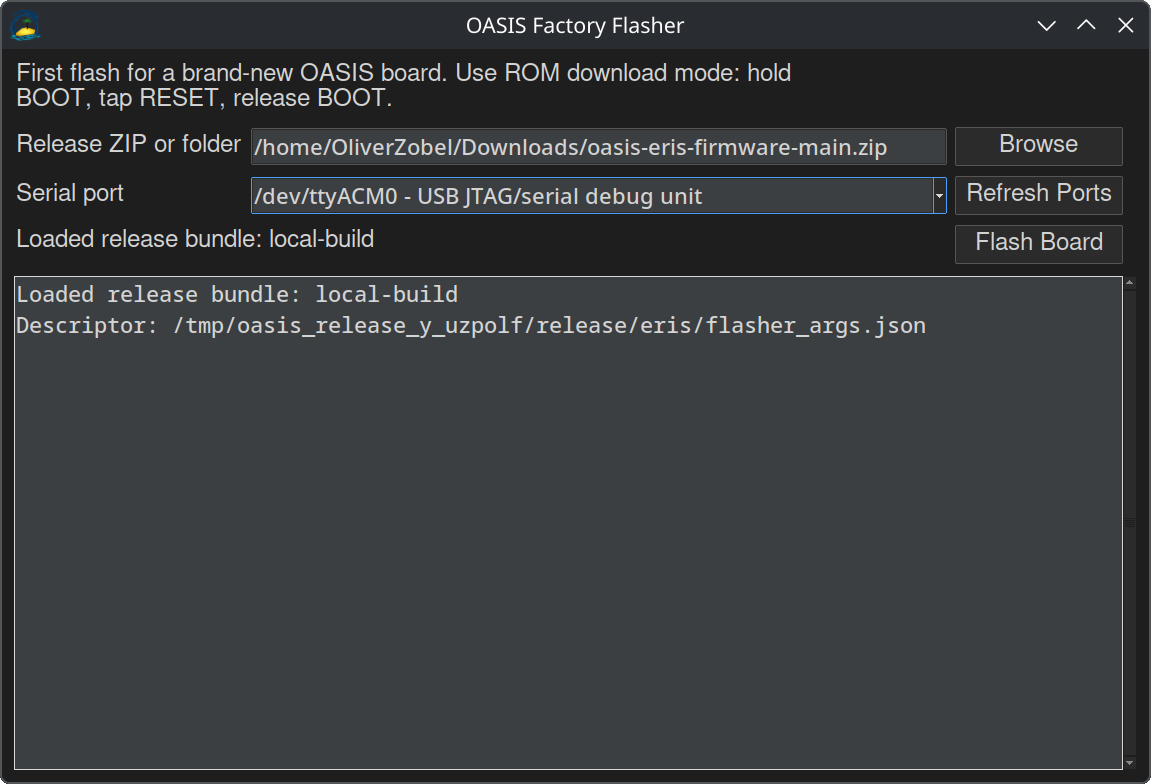}
	\end{subfigure}
	\caption{Screenshots of the \textit{OASIS-FactoryFlasher} on all three supported operating systems. The current version can be downloaded from the \textit{GitLab} page: \url{https://gitlab.com/oasis-acquisition/oasis-factoryflasher}.}
	\label{fig:flasher}
\end{figure}

To flash the bootloader, configure the \textit{ESP32-S3} options, and upload the \textit{OASIS-Firmware}, the \textit{OASIS-FactoryFlasher} is provided. This is a \textit{Python} GUI application that automatically performs the aforementioned steps. The \textit{OASIS-FactoryFlasher} is available and tested for \textit{Windows}, \textit{MacOS}, and \textit{Linux} from the \textit{GitLab} page: \url{https://gitlab.com/oasis-acquisition/oasis-factoryflasher}; see also \cref{fig:flasher}.

\vspace{-0.4cm}

To set up a new \textit{OASIS-ERIS} board (or also an \textit{OASIS-UROS} board), follow these instructions:\vspace{-0.4cm}
\begin{enumerate}
	\item Download the pre-built binary of the \textit{OASIS-FactoryFlasher} for your operating system (\url{https://gitlab.com/oasis-acquisition/oasis-factoryflasher}).\vspace{-0.2cm}
	\item Download the pre-built firmware from the OASIS-Firmware repository (\url{https://gitlab.com/oasis-acquisition/oasis-firmware/-/releases}).\vspace{-0.2cm}
	\item Launch the \textit{OASIS-FactoryFlasher} executable. Note: On \textit{Linux}, you need superuser privileges to access the serial device.\vspace{-0.2cm}
	\item Hold down the 'BOOT' button on the \textit{OASIS} board, connect it to your device, and then release the button after connection. Alternatively, hold down the 'BOOT' button and then press the 'RESET' button.\vspace{-0.2cm}
	\item Select the downloaded firmware zip-archive in the \textit{OASIS-FactoryFlasher} using 'Browse'.\vspace{-0.2cm}
	\item Select the \textit{OASIS} board under 'Serial port'. Typical port names:\vspace{-0.1cm}
	\begin{itemize}
		\item Windows: COM9 - USB Serial Device (COM9)
		\item macOS: /dev/cu.usbmodem1101 - USB JTAG/serial debug unit
		\item Linux: /dev/ttyACM0 - USB JTAG/serial debug unit
	\end{itemize}
	\item Click 'Flash Board'.\vspace{-0.2cm}
	\item Press the 'RESET' button once to reset \textit{OASIS} and load the firmware.\vspace{-0.2cm}
\end{enumerate}\vspace{-0.4cm}

After this, the \textit{OASIS-Firmware} should boot, as indicated by a LED effect and sound. If the board is connected to a PC/laptop and an SD card is inserted, the SD card should appear as a mass storage device in the operating system.

\subsubsection{Setting device metadata}

Lastly, it is recommended to set the device metadata. This includes a unique name for the board, the architecture (\textit{OASIS-ERIS} or \textit{OASIS-UROS}), hardware version, ADC resolution, and additional device features (currently unused). Most easily, this is done using the \textit{OASIS-GUI} or \textit{OASIS-TUI} and following the on-screen instruction; the setup of these components is shown in the next section.
% !TeX spellcheck = en_US
\section{Operation instructions}\label{sec:OperatingInstructions}

Communication with \textit{OASIS} boards is possible via a serial interface using SCPI (Standard Commands for Programmable Instruments) or custom commands; see the command reference in \cref{sec:CommandRef} for more details. For ease of use, a couple of Python packages are provided to handle the communication and provide a high-level user interface. All packages were tested and validated under Windows, MacOS, and Linux. The basis is the \textit{OASIS-API}, which implements the communication with \textit{OASIS} boards and handles the start, retrieval, and post-processing of sample data. Built on top of the \textit{OASIS-API} are the \textit{OASIS-TUI}, a terminal user interface, and the \textit{OASIS-GUI}, a graphical user interface. The installation and operation of the Python components for \textit{OASIS} boards are detailed below. Lastly, the firmware update process is explained.

\subsection{Installation of \textit{Python} components}\label{sec:python}

First, \textit{Python} must be installed on the system by following the instructions for the operating system of choice on the \textit{Python} website\footnote{\url{https://www.python.org/downloads/}}. Next, a virtual environment\footnote{\url{https://docs.python.org/3/library/venv.html}} should be created and activated to separate the install from the global \textit{Python} installation and other projects. The \textit{OASIS} packages can then be installed from the \textit{Python Package Index}\footnote{\url{https://pypi.org/}} (PyPI) using the \textit{Package Installer for Python} (pip):\vspace{-0.4cm}
\begin{itemize}
	\item Installation of \textit{OASIS-GUI}\footnote{\url{https://gitlab.com/oasis-acquisition/oasis-gui}}: \texttt{pip install oasis-gui}\vspace{-0.2cm}
	\item Installation of \textit{OASIS-TUI}\footnote{\url{https://gitlab.com/oasis-acquisition/oasis-tui}}: \texttt{pip install oasis-tui}\vspace{-0.2cm}
	\item Installation of \textit{OASIS-API}\footnote{\url{https://gitlab.com/oasis-acquisition/oasis-api}}: \texttt{pip install oasis-api}
\end{itemize}\vspace{-0.4cm}
The respective installation command automatically installs all required dependencies; for instance, when installing the \textit{OASIS-GUI} or \textit{OASIS-TUI}, the \textit{OASIS-API} is installed automatically.

All \textit{Python} packages and the \textit{OASIS-Firmware} follow a three-digit versioning scheme, i.e., the version number is X.Y.Z, where X is the major release, Y is the minor release, and Z is the bug fix release. For everything to function as intended, the major and minor releases have to match.

\subsection{Using the \textit{OASIS-GUI}}\label{sec:gui}

\begin{figure}[H]
	\includegraphics[width=\textwidth]{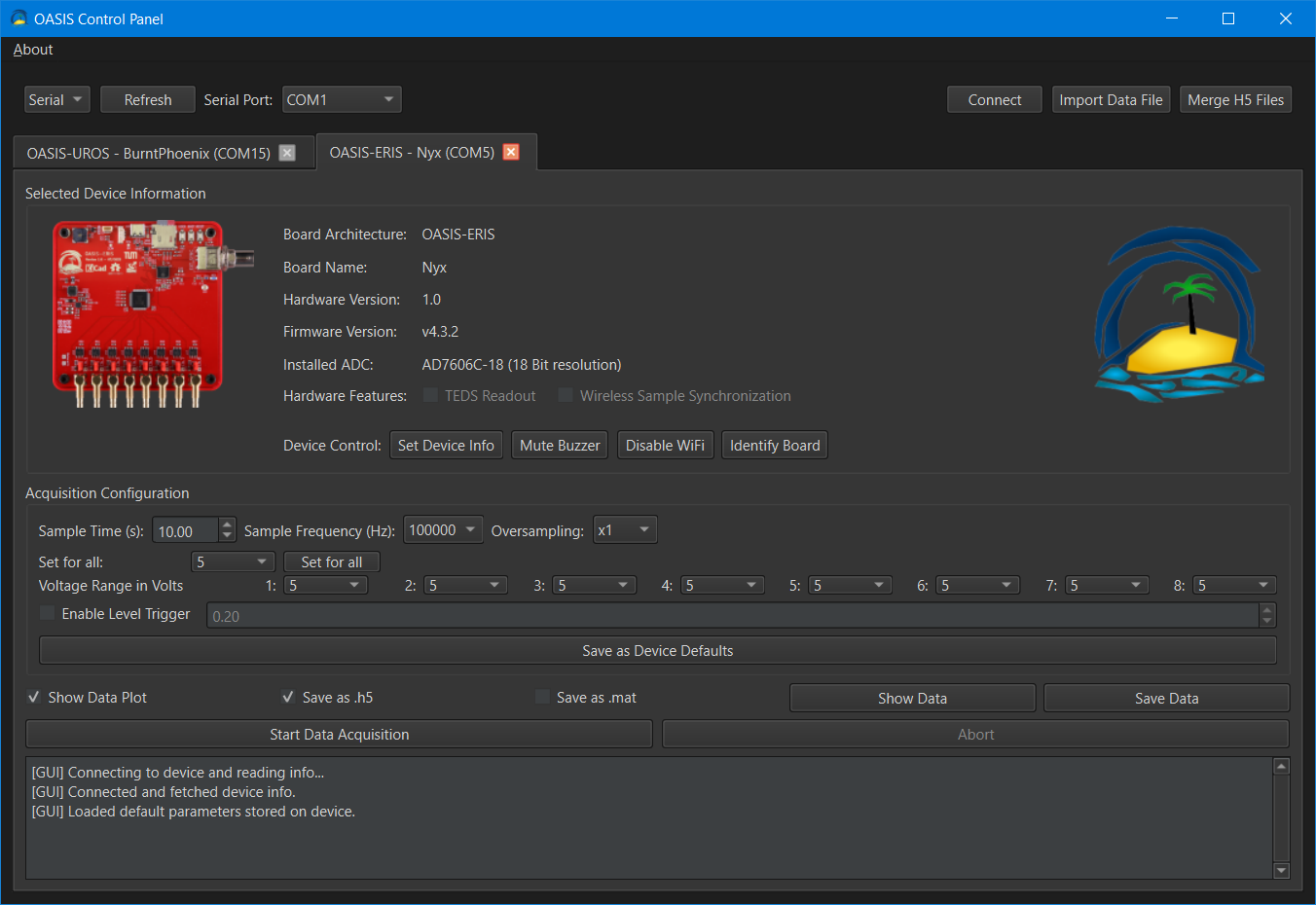}
	\caption{Screenshot of the \textit{OASIS-GUI} with two connected boards.}
	\label{fig:oasis_gui}
\end{figure}

The \textit{OASIS-GUI} can be opened from a terminal window with an activated virtual environment using the command:\vspace{-0.4cm}
\begin{itemize}
	\item[\ding{229}] \texttt{oasis-gui}
\end{itemize}\vspace{-0.4cm}
A screenshot of the \textit{OASIS-GUI} can be seen in \cref{fig:oasis_gui}.

\vspace{-0.4cm}

Using the buttons in the top, \textit{OASIS} boards can be connected to via serial communication or TCP. Each connected board is then displayed as a tab in the GUI. For each device, the information stored in the device metadata, e.g., the name and firmware version, is displayed. Below are the settings for data acquisition. The sample time can be freely selected, while the sampling frequency is limited to a few options. This is because, at arbitrary frequencies, the PWM timer cannot achieve the exact frequency due to limited timer resolution, resulting in an erroneous sampling frequency. 

Additionally, the oversampling integrated in the ADC can be enabled. Oversampling means that the ADC acquires multiple samples and averages them for each request one. This reduces the amount of noise in the acquired signals. The available oversampling factors are summarized in \cref{tab:OSOASIS}. Further, the voltage range of each channel can be configured individually from the available options summarized in \cref{tab:VROASIS}. Lastly, a level trigger active on the first channel can be enabled, and the trigger voltage configured. It is also possible to save the configured device parameters as the default parameter set for each device. This is especially required when multiple boards are synchronized.

\begin{table}[h!]
	\centering
	\begin{tabular}{c|c|c|c|c|c|c|c|c|c} 
		Oversampling factor & x1 (Off) & x2 & x4 & x8 & x16 & x32 & x64 & x128 & x256 \\\hline
		Theoretical maximum throughput (kSPS) & 1000 & 500 & 250 & 125 & 62.5 & 31.25 & 15.6 & 7.8 & 3.9
	\end{tabular}
	\caption{Oversampling factors and maximum theoretical throughput in kilo samples per second (kSPS)~\cite{ADCDatasheet}.}
	\label{tab:OSOASIS}
\end{table}

\begin{table}[h!]
	\centering
	\begin{tabular}{c|c|c|c|c|c} 
		Voltage range  & $\pm\SI{2.5}{\volt}$ & $\pm\SI{5}{\volt}$ & $\pm\SI{6.25}{\volt}$ & $\pm\SI{10}{\volt}$ & $\pm\SI{12.5}{\volt}$ \\\hline
		Voltage resolution & \SI{19}{\micro\volt} & \SI{38.1}{\micro\volt} & \SI{47.7}{\micro\volt} & \SI{76.3}{\micro\volt} & \SI{95.36}{\micro\volt}
	\end{tabular}
	\caption{Available voltage ranges and corresponding resolution~\cite{ADCDatasheet}.}
	\label{tab:VROASIS}
\end{table}

Below the acquisition parameters, there are options for handling the sampled data after acquisition. The checkbox options on the left allow for automatic plotting or saving the data after the data acquisition. Using the buttons on the right, it is possible to manually plot or save the previously acquired data.

The data acquisition is started with the 'Start Data Acquisition' button just above the log output. When multiple boards are connected, the \textit{OASIS-GUI} automatically fetches the files from the additional boards, includes them in the data plot, and saves a single combined file containing all sampled channels.

\subsection{Using the \textit{OASIS-TUI}}\label{sec:tui}

\begin{figure}[H]
	\centering
	\includegraphics[width=0.815\textwidth]{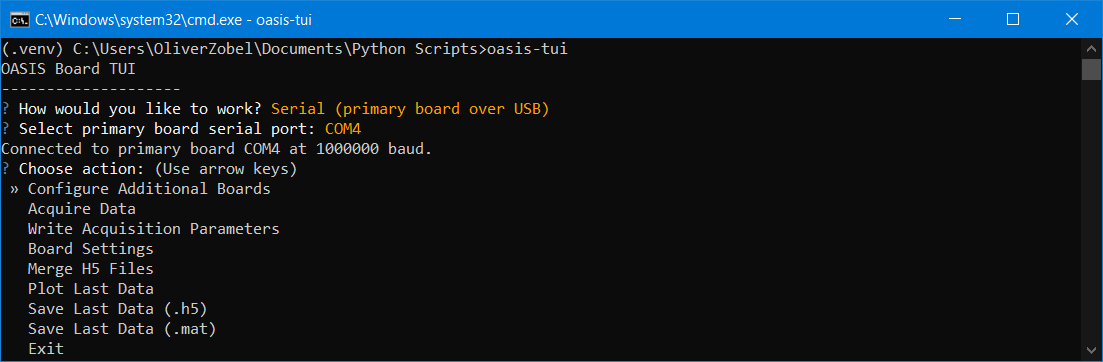}
	\caption{Screenshot of a started \textit{OASIS-TUI} session with one connected board.}
	\label{fig:oasis_tui}
\end{figure}

The \textit{OASIS-TUI} offers the same features but without a graphical user interface. Instead, a guided terminal user interface is provided, which is especially suited for remote operation, for example, over SSH. The \textit{OASIS-TUI}, see also \cref{fig:oasis_tui}, can be opened from a terminal window with an activated virtual environment using the command:\vspace{-0.4cm}
\begin{itemize}
	\item[\ding{229}] \texttt{oasis-tui}
\end{itemize}\vspace{-0.4cm}

\vspace{-0.4cm}

As with the \textit{OASIS-GUI}, \textit{OASIS} boards can be connected to using a serial or TCP connection. Additional boards can be added using 'Configure Additional Boards'. The \textit{OASIS-TUI} then automatically retrieves the sampled data from the additional boards and merges them into a single file.

\subsection{Custom scripting using the \textit{OASIS-API}}\label{sec:api}

The greatest degree of customization is possible with the \textit{OASIS-API}, which provides high-level commands that can be integrated into custom \textit{Python} scripts. For example, in a staged measurement script that first activates a test rig, applies an input, and then, after some settling time, acquires data using \textit{OASIS}. An example \textit{Python} script for data acquisition using the \textit{OASIS-API} might look like this:

\lstset{style=Python}
\begin{lstlisting}[label={lst:pyModal1},caption={Import of Python packages, definition of measurement data and calibration factor.}]
import os
from oasis_api import OasisBoard

serial_port = os.getenv("OASIS_SERIAL_PORT", "/dev/cu.usbmodem11101")
baudrate = int(os.getenv("OASIS_SERIAL_BAUDRATE", "1000000"))

board = OasisBoard(
    mode="serial",
    port=serial_port,
    baudrate=baudrate,
)

board.connect()
board.set_parameters(
    t_sample=1.0,
    f_sample=10000,
    voltage_range=[2.5] * 8,
    trigger=False,
    v_trigg=0.5,
    oversampling=2,
    sync_mode=0,
)

board.acquire()
board.save_data_h5("mydata.h5", overwrite=True)
\end{lstlisting}\vspace{-0.2cm}

More examples can be found on \textit{GitLab}: \url{https://gitlab.com/oasis-acquisition/oasis-api/-/tree/main/examples}.

\subsection{Firmware update}

The updated \textit{OASIS-Firmware} can either be compiled locally or the pre-compiled binaries can be downloaded form the \textit{GitLab} release page: \url{https://gitlab.com/oasis-acquisition/oasis-firmware/-/releases}. After the firmware for the correct board, e.g., for \textit{OASIS-ERIS}, was downloaded, follow these steps:\vspace{-0.4cm}
\begin{enumerate}
	\item Extract the file \texttt{firmware.bin} from the downloaded zip archive (or use a locally compiled one).\vspace{-0.2cm}
	\item Connect the \textit{OASIS} board to the PC/laptop with a USB-C cable.\vspace{-0.2cm}
	\item Locate the SD card mounted through the \textit{OASIS} board in your desktop environment.\vspace{-0.2cm}
	\item Copy the file \texttt{firmware.bin} to the SD card.\vspace{-0.2cm}
	\item Press 'RESET' on the \textit{OASIS} board. The board now reboots and applies the new firmware.
\end{enumerate}\vspace{-0.4cm}
% !TeX spellcheck = en_US

\section{Validation and characterization}

In this section, the hardware and software of \textit{OASIS-ERIS} are validated. First, the inter-device synchronization is tested using a signal generator for normal and triggered data acquisition. Second, the performance of \textit{OASIS-ERIS} for structural dynamic measurements is compared with a commercial system using Frequency Response Function (FRF) measurements and Experimental Modal Analysis. The raw time data acquired for the validation of \textit{OASIS-ERIS} is available for download at~\cite{OASISERISData}.

\subsection{Validation of inter-device synchronicity}

\begin{figure}[H]
	\centering
	\scriptsize
	\def\svgwidth{0.8\textwidth}
	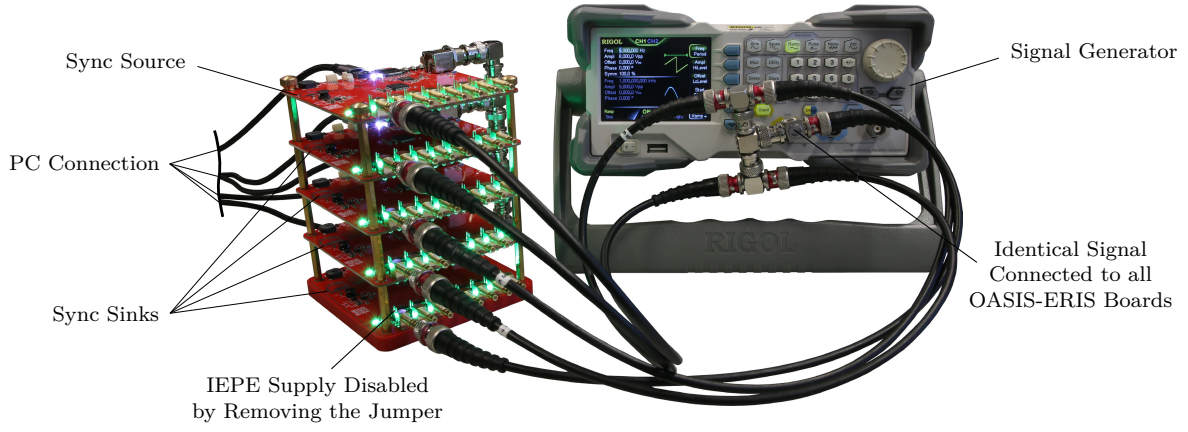
	\caption{Overview of inter-device synchronicity test with a \textit{RIGOL DG1032Z} signal generator and five \textit{OASIS-ERIS} boards.}
	\label{fig:sync_test}
\end{figure}

To validate the inter-device synchronization, the signal from a \textit{RIGOL DG1032Z} signal generator is connected to all first channels of five \textit{OASIS-ERIS} boards; see also \cref{fig:sync_test}. For the tests, a ramp waveform with \SI{100}{\percent} symmetry, a frequency of \SI{5}{\hertz}, and an amplitude of $\pm\SI{4}{\volt}$ is used. The IEPE supply of the \textit{OASIS-ERIS} boards is disabled for these tests by removing the jumpers near the SMB connector. Data is acquired with a sampling frequency of \SI{100}{\kilo\hertz} for a duration of \SI{10}{\second} for the normal sample, and for a duration of \SI{1}{\second} for the triggered sampling. For each configuration, three measurements are performed.

To evaluate the time delay between the \textit{Sync Source} and the attached \textit{Sync Sinks}, the cross-correlation is used. Since all \textit{OASIS-ERIS} boards are fed the same signal, the point where the cross-correlation is maximal denotes when the signals are synchronized. This is evaluated below for both normal and triggered sampling.

\vspace{-0.4cm}

\subsubsection{Synchronicity of normal sample}

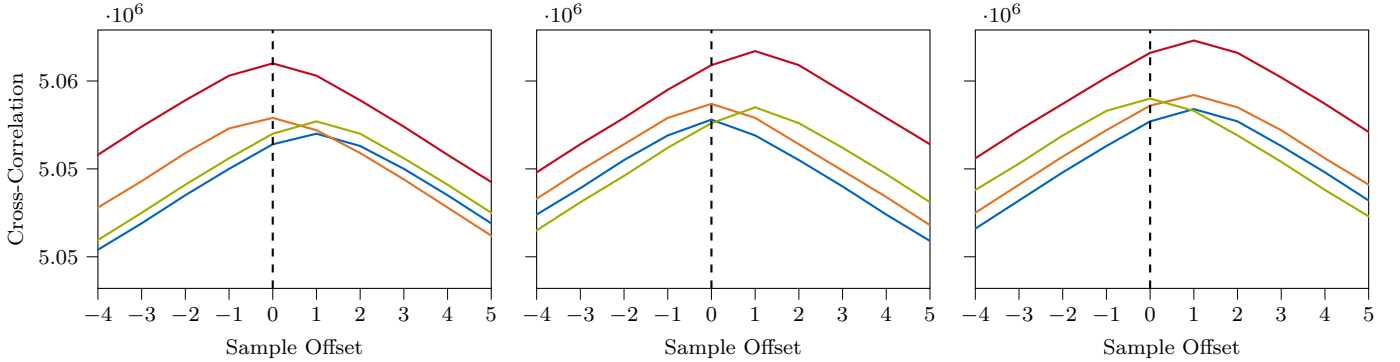
\begin{figure}[H]
	\centering
	\scriptsize
	% This file was created with matplot2tikz v0.5.4.
\begin{tikzpicture}

\definecolor{darkgray176}{RGB}{176,176,176}

\begin{groupplot}[group style={group size=3 by 1, horizontal sep=0.6cm}]
\nextgroupplot[
height=5.0cm,
tick align=outside,
tick pos=left,
width=0.365\textwidth,
x grid style={darkgray176},
xlabel={Sample Offset},
xmin=-4, xmax=5,
xtick style={color=black},
xtick={-5,-4,-3,-2,-1,0,1,2,3,4,5,6},
y grid style={darkgray176},
ylabel={Cross-Correlation},
ymin=5.0482e+06, ymax=5.0629e+06,
ytick style={color=black},
]
\addplot [thick, TUMBlue]
table {%
-5 5.0488e+06
-4 5.0504e+06
-3 5.0519e+06
-2 5.0535e+06
-1 5.055e+06
0 5.0564e+06
1 5.057e+06
2 5.0563e+06
3 5.055e+06
4 5.0535e+06
5 5.0519e+06
6 5.0504e+06
};
\addplot [thick, TUMOrange]
table {%
-5 5.0512e+06
-4 5.0528e+06
-3 5.0543e+06
-2 5.0559e+06
-1 5.0573e+06
0 5.0579e+06
1 5.0572e+06
2 5.0559e+06
3 5.0544e+06
4 5.0528e+06
5 5.0512e+06
6 5.0497e+06
};
\addplot [thick, TUMGreen]
table {%
-5 5.0494e+06
-3 5.0525e+06
-2 5.0541e+06
-1 5.0556e+06
0 5.057e+06
1 5.0577e+06
2 5.057e+06
3 5.0556e+06
4 5.0541e+06
5 5.0525e+06
6 5.051e+06
};
\addplot [thick, TUMDiagRed1]
table {%
-5 5.0543e+06
-4 5.0558e+06
-3 5.0574e+06
-2 5.0589e+06
-1 5.0603e+06
0 5.061e+06
1 5.0603e+06
2 5.0589e+06
3 5.0574e+06
4 5.0558e+06
6 5.0527e+06
};
\addplot [thick, black, dashed]
table {%
0 5.0482e+06
0 5.0629e+06
};

\nextgroupplot[
height=5.0cm,
tick align=outside,
tick pos=left,
width=0.365\textwidth,
x grid style={darkgray176},
xlabel={Sample Offset},
xmin=-4, xmax=5,
xtick style={color=black},
xtick={-5,-4,-3,-2,-1,0,1,2,3,4,5,6},
y grid style={darkgray176},
ymin=5.0482e+06, ymax=5.0629e+06,
ytick style={color=black},
yticklabels={}
]
\addplot [thick, TUMBlue]
table {%
-5 5.0508e+06
-4 5.0524e+06
-3 5.0539e+06
-2 5.0555e+06
-1 5.0569e+06
0 5.0578e+06
1 5.0569e+06
2 5.0555e+06
3 5.054e+06
4 5.0524e+06
5 5.0509e+06
6 5.0493e+06
};
\addplot [thick, TUMOrange]
table {%
-5 5.0518e+06
-4 5.0533e+06
-3 5.0549e+06
-2 5.0564e+06
-1 5.0579e+06
0 5.0587e+06
1 5.0579e+06
2 5.0564e+06
3 5.0549e+06
4 5.0534e+06
5 5.0518e+06
6 5.0502e+06
};
\addplot [thick, TUMGreen]
table {%
-5 5.0499e+06
-4 5.0515e+06
-3 5.0531e+06
-2 5.0546e+06
-1 5.0562e+06
0 5.0576e+06
1 5.0585e+06
2 5.0576e+06
3 5.0562e+06
4 5.0547e+06
5 5.0531e+06
6 5.0516e+06
};
\addplot [thick, TUMDiagRed1]
table {%
-5 5.0533e+06
-4 5.0548e+06
-3 5.0564e+06
-2 5.0579e+06
-1 5.0595e+06
0 5.0609e+06
1 5.0617e+06
2 5.0609e+06
3 5.0594e+06
4 5.0579e+06
5 5.0564e+06
6 5.0548e+06
};
\addplot [thick, black, dashed]
table {%
0 5.0482e+06
0 5.0629e+06
};

\nextgroupplot[
height=5.0cm,
tick align=outside,
tick pos=left,
width=0.365\textwidth,
x grid style={darkgray176},
xlabel={Sample Offset},
xmin=-4, xmax=5,
xtick style={color=black},
xtick={-5,-4,-3,-2,-1,0,1,2,3,4,5,6},
y grid style={darkgray176},
ymin=5.0482e+06, ymax=5.0629e+06,
ytick style={color=black},
yticklabels={}
]
\addplot [thick, TUMBlue]
table {%
-5 5.05e+06
-4 5.0516e+06
-3 5.0532e+06
-2 5.0548e+06
-1 5.0563e+06
0 5.0577e+06
1 5.0584e+06
2 5.0577e+06
3 5.0563e+06
4 5.0548e+06
5 5.0532e+06
6 5.0517e+06
};
\addplot [thick, TUMOrange]
table {%
-5 5.0509e+06
-4 5.0525e+06
-3 5.0541e+06
-2 5.0557e+06
-1 5.0572e+06
0 5.0586e+06
1 5.0592e+06
2 5.0585e+06
3 5.0572e+06
4 5.0556e+06
5 5.0541e+06
6 5.0525e+06
};
\addplot [thick, TUMGreen]
table {%
-5 5.0522e+06
-4 5.0538e+06
-3 5.0553e+06
-2 5.0569e+06
-1 5.0583e+06
0 5.059e+06
1 5.0583e+06
2 5.0569e+06
3 5.0554e+06
4 5.0538e+06
5 5.0523e+06
6 5.0507e+06
};
\addplot [thick, TUMDiagRed1]
table {%
-5 5.054e+06
-4 5.0556e+06
-3 5.0572e+06
-2 5.0587e+06
-1 5.0602e+06
0 5.0616e+06
1 5.0623e+06
2 5.0616e+06
3 5.0602e+06
4 5.0587e+06
5 5.0571e+06
6 5.0555e+06
};
\addplot [thick, black, dashed]
table {%
0 5.0482e+06
0 5.0629e+06
};
\end{groupplot}
\end{tikzpicture}
	\caption{Results of synchronization test for \textbf{normal sampling} showing the normalized cross-correlation between the \textit{Sync Source} and the four \textit{Sync Sinks} for all three performed measurements.}
	\label{fig:sync_normal}
\end{figure}

The results for normal sampling are shown in \cref{fig:sync_normal}. As shown, the maximum sample offset across all boards and tests is 1 sample. For a sample rate of \SI{100}{\kilo\hertz}, this corresponds to a maximum delay of \SI{10}{\micro\second}; however, this is the worst case. Across the three tests, the sample offset ranges from 0 to 1 for all boards.

\vspace{-0.4cm}

\subsubsection{Synchronicity of triggered sample}

For the triggered sampling, all boards are always perfectly synchronized; see also \cref{fig:sync_trigger}. This difference could be explained by the different sampling approaches: while normal sampling starts immediately, triggered sampling discards some data before the trigger. In particular, the first \textit{CONVST} pulse might be missed due to latency in software scheduling. Since triggered sampling discards the beginning of the continuously sampled data, this issue cannot be observed with triggered sampling.

\begin{figure}[H]
	\centering
	\scriptsize
	% This file was created with matplot2tikz v0.5.4.
\begin{tikzpicture}

\definecolor{darkgray176}{RGB}{176,176,176}

\begin{groupplot}[group style={group size=3 by 1, horizontal sep=1.2cm}]
\nextgroupplot[
height=5.0cm,
tick align=outside,
tick pos=left,
width=0.346\textwidth,
x grid style={darkgray176},
xlabel={Sample Offset},
xmin=-4, xmax=5,
xtick style={color=black},
xtick={-5,-4,-3,-2,-1,0,1,2,3,4,5,6},
y grid style={darkgray176},
ylabel={Cross-Correlation},
ymin=5.0287e+05, ymax=5.0426e+05,
ytick style={color=black}
]
\addplot [thick, TUMBlue]
table {%
-5 5.0309e+05
-4 5.0324e+05
-3 5.034e+05
-2 5.0356e+05
-1 5.0369e+05
0 5.0376e+05
1 5.0369e+05
2 5.0356e+05
3 5.034e+05
4 5.0325e+05
5 5.0309e+05
6 5.0294e+05
};
\addplot [thick, TUMOrange]
table {%
-5 5.032e+05
-4 5.0335e+05
-3 5.0351e+05
-2 5.0366e+05
-1 5.038e+05
0 5.0386e+05
1 5.038e+05
2 5.0366e+05
3 5.0351e+05
4 5.0335e+05
5 5.032e+05
6 5.0304e+05
};
\addplot [thick, TUMGreen]
table {%
-5 5.0317e+05
-4 5.0333e+05
-3 5.0348e+05
-2 5.0364e+05
-1 5.0378e+05
0 5.0384e+05
1 5.0378e+05
2 5.0364e+05
3 5.0349e+05
4 5.0333e+05
5 5.0317e+05
6 5.0302e+05
};
\addplot [thick, TUMDiagRed1]
table {%
-5 5.0353e+05
-4 5.0369e+05
-3 5.0384e+05
-2 5.0399e+05
-1 5.0413e+05
0 5.0419e+05
1 5.0413e+05
2 5.0399e+05
3 5.0384e+05
4 5.0368e+05
5 5.0353e+05
6 5.0337e+05
};
\addplot [thick, black, dashed]
table {%
0 5.0287e+05
0 5.0426e+05
};

\nextgroupplot[
height=5.0cm,
tick align=outside,
tick pos=left,
width=0.346\textwidth,
x grid style={darkgray176},
xlabel={Sample Offset},
xmin=-4, xmax=5,
xtick style={color=black},
xtick={-5,-4,-3,-2,-1,0,1,2,3,4,5,6},
y grid style={darkgray176},
ymin=5.0449e+05, ymax=5.0588e+05,
ytick style={color=black}
]
\addplot [thick, TUMBlue]
table {%
-5 5.047e+05
-4 5.0486e+05
-3 5.0502e+05
-2 5.0517e+05
-1 5.0531e+05
0 5.0537e+05
1 5.0531e+05
2 5.0517e+05
3 5.0502e+05
4 5.0486e+05
5 5.0471e+05
6 5.0455e+05
};
\addplot [thick, TUMOrange]
table {%
-5 5.0482e+05
-4 5.0497e+05
-3 5.0513e+05
-2 5.0528e+05
-1 5.0542e+05
0 5.0548e+05
1 5.0542e+05
2 5.0528e+05
3 5.0513e+05
4 5.0497e+05
5 5.0482e+05
6 5.0466e+05
};
\addplot [thick, TUMGreen]
table {%
-5 5.0479e+05
-4 5.0494e+05
-3 5.051e+05
-2 5.0525e+05
-1 5.0539e+05
0 5.0546e+05
1 5.0539e+05
2 5.0526e+05
3 5.051e+05
4 5.0495e+05
5 5.0479e+05
6 5.0464e+05
};
\addplot [thick, TUMDiagRed1]
table {%
-5 5.0515e+05
-4 5.0531e+05
-3 5.0546e+05
-2 5.0562e+05
-1 5.0575e+05
0 5.0582e+05
1 5.0575e+05
2 5.0561e+05
3 5.0546e+05
4 5.053e+05
5 5.0515e+05
6 5.0499e+05
};
\addplot [thick, black, dashed]
table {%
0 5.0449e+05
0 5.0588e+05
};

\nextgroupplot[
height=5.0cm,
tick align=outside,
tick pos=left,
width=0.346\textwidth,
x grid style={darkgray176},
xlabel={Sample Offset},
xmin=-4, xmax=5,
xtick style={color=black},
xtick={-5,-4,-3,-2,-1,0,1,2,3,4,5,6},
y grid style={darkgray176},
ymin=5.0449e+05, ymax=5.0588e+05,
ytick style={color=black}
]
\addplot [thick, TUMBlue]
table {%
-5 5.047e+05
-4 5.0486e+05
-3 5.0502e+05
-2 5.0517e+05
-1 5.0531e+05
0 5.0537e+05
1 5.0531e+05
2 5.0517e+05
3 5.0502e+05
4 5.0486e+05
5 5.0471e+05
6 5.0455e+05
};
\addplot [thick, TUMOrange]
table {%
-5 5.0482e+05
-4 5.0497e+05
-3 5.0513e+05
-2 5.0528e+05
-1 5.0542e+05
0 5.0548e+05
1 5.0542e+05
2 5.0528e+05
3 5.0513e+05
4 5.0497e+05
5 5.0482e+05
6 5.0466e+05
};
\addplot [thick, TUMGreen]
table {%
-5 5.0479e+05
-4 5.0494e+05
-3 5.051e+05
-2 5.0525e+05
-1 5.0539e+05
0 5.0546e+05
1 5.0539e+05
2 5.0526e+05
3 5.051e+05
4 5.0495e+05
5 5.0479e+05
6 5.0464e+05
};
\addplot [thick, TUMDiagRed1]
table {%
-5 5.0515e+05
-4 5.0531e+05
-3 5.0546e+05
-2 5.0562e+05
-1 5.0575e+05
0 5.0582e+05
1 5.0575e+05
2 5.0561e+05
3 5.0546e+05
4 5.053e+05
5 5.0515e+05
6 5.0499e+05
};
\addplot [thick, black, dashed]
table {%
0 5.0449e+05
0 5.0588e+05
};
\end{groupplot}
\end{tikzpicture}
	\caption{Results of synchronization test for \textbf{triggered sampling} showing the normalized cross-correlation between the \textit{Sync Source} and the four \textit{Sync Sinks} for all three performed measurements.}
	\label{fig:sync_trigger}
\end{figure}
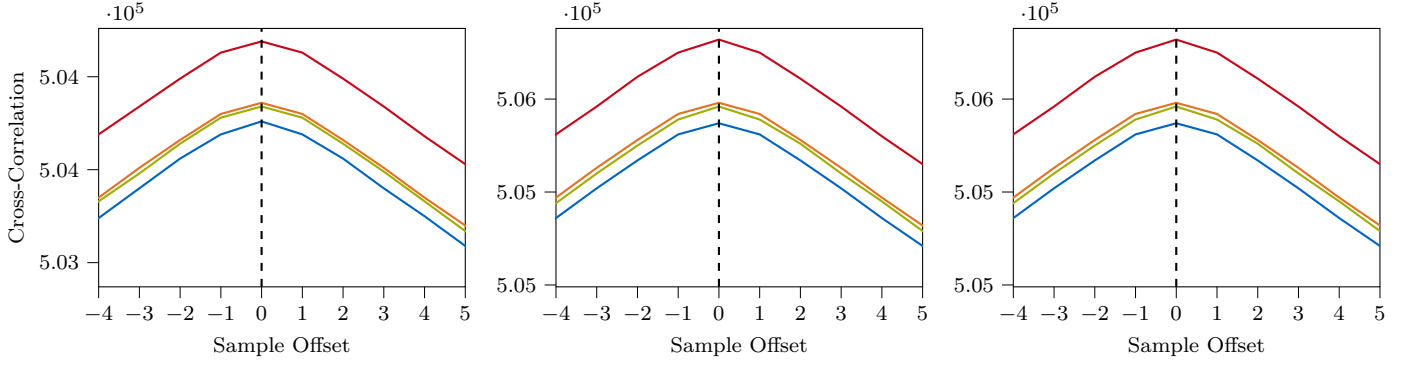\vspace{-0.4cm}

\subsection{Structural dynamic validation using experimental modal analysis}

To evaluate the performance of \textit{OASIS-ERIS} for structural dynamics measurements, an experimental modal analysis campaign is performed. For this, Frequency Response Functions (FRFs) are measured using shaker testing, once with \textit{OASIS-ERIS} boards and once with a commercial measurement system. After this, modal parameters are identified for both systems using \textit{pyFBS}~\cite{pyFBS}.

\subsubsection{Experimental test setup}

\begin{figure}[H]
	\centering
	\scriptsize
	\def\svgwidth{\textwidth}
	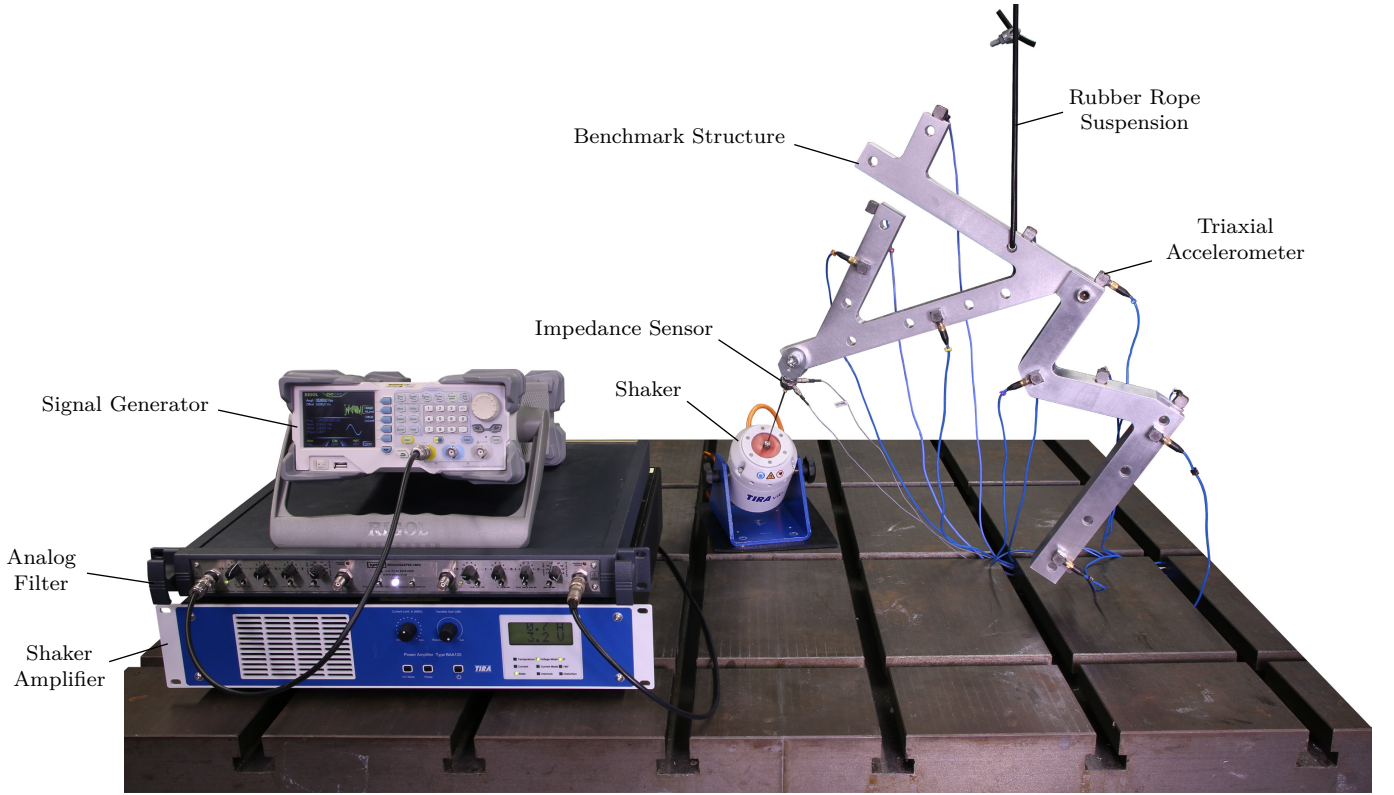
	\caption{Overview of the Experimental Modal Analysis test setup with shaker excitation.}
	\label{fig:ema_setup}
\end{figure}

An overview of the measurement setup is shown in \cref{fig:ema_setup}.

\vspace{-0.4cm}

\newpage

The Modal Analysis is performed on an aluminum benchmark structure, commonly used for substructuring, that is suspended from a rubber rope. To excite the system, a \textit{TIRAvib S 50018} shaker is used. The shaker is connected to the system through a stinger and a \textit{Dytran 5860B} impedance sensor.

As an excitation signal, random noise generated by a \textit{RIGOL DG1032Z} signal generator is used. Since the noise generated by the signal generator has a bandwidth of \SI{30}{\mega\hertz}, a \textit{KEMO Benchmaster VBF8} analog filter is used to limit the bandwidth and amplify the signal. In the first filter stage, an eight-pole Butterworth low-pass filter with a cut-off frequency of \SI{5}{\kilo\hertz} is applied. In the second stage, an eight-pole Butterworth high-pass filter with a cut-off frequency of \SI{2}{\hertz} is applied. The second filter is applied to remove low-frequency vibrations with large displacement amplitudes.

\begin{figure}[H]
	\centering
	\scriptsize
	\def\svgwidth{0.44\textwidth}
	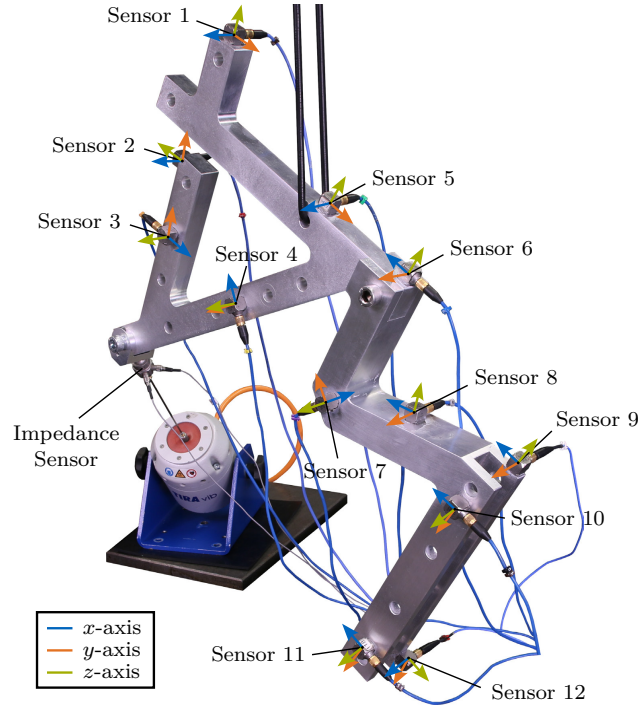
	\caption{Overview of the Experimental Modal Analysis test setup with shaker excitation.}
	\label{fig:ema_sensors}
\end{figure}

The structure's response is measured using 12 \textit{Kistler 8688A50} triaxial accelerometers glued to the structure.  An overview of the sensor setup is given in \cref{fig:ema_sensors}.

\vspace{-0.4cm}

 Measurements are performed using five \textit{OASIS-ERIS} boards and a \textit{Müller-BBM PAK MKII} system with the acquisition parameters summarized in \cref{tab:ema_para}. The channel assignment for the \textit{OASIS-ERIS} boards is given in \cref{tab:ema_oasis_ch}. For the \textit{Müller-BBM PAK MKII} system, the same sensor channel order is used.

\begin{table}[H]
	\centering
	\caption{Summary of Experimental Modal Analysis acquisition parameters.}
	\label{tab:ema_para}
	\begin{tabular}{c|c|c}
		& \textit{OASIS-ERIS} & \textit{Müller-BBM PAK MKII} \\\hline
		Sampling Frequency & \SI{16}{\kilo\hertz} & \SI{16.384}{\kilo\hertz} \\\hline
		Measurement Duration & \multicolumn{2}{c}{\SI{10}{\second}} \\\hline
		Voltage Range & $\pm\SI{5}{\volt}$ & $\pm\SI{10}{\volt}$ \\\hline
		Oversampling Factor & x32 & Unknown \\\hline
		Number of Measurements & \multicolumn{2}{c}{20}
	\end{tabular}
\end{table}

\begin{table}[H]
	\centering
	\caption{Channel assignment of \textit{OASIS-ERIS} boards for Experimental Modal Analysis setup.}
	\label{tab:ema_oasis_ch}
	\begin{tabular}{c|c|c|c|c|c|c|c|c}
		Board & Ch. 1 & Ch. 2 & Ch. 3 & Ch. 4 & Ch. 5 & Ch. 6 & Ch. 7 & Ch. 8 \\\hline
		\textit{OASIS-ERIS} ALPHA & Imp. Force & Imp. Acc. & S1+X & S1+Y & S1+Z & S2+X & S2+Y & S2+Z \\\hline
		\textit{OASIS-ERIS} BETA & S3+X & S3+Y & S3+Z & S4+X & S4+Y & S4+Z & S5+X & S5+Y \\\hline
		\textit{OASIS-ERIS} GAMMA & S5+Z & S6+X & S6+Y & S6+Z & S7+X & S7+Y & S7+Z & S8+X \\\hline
		\textit{OASIS-ERIS} DELTA & S8+Y & S8+Z & S9+X & S9+Y & S9+Z & S10+X & S10+Y & S10+Z \\\hline
		\textit{OASIS-ERIS} EPSILON & S11+X & S11+Y & S11+Z & S12+X & S12+Y & S12+Z & &
	\end{tabular}
\end{table}

\subsubsection{Frequency response function measurements}

For both systems, the FRFs are estimated using \textit{pyFRF}~\cite{pyFRF}. To reduce leakage, a Hanning window is applied to the excitation and the responses. Since the signal level of the excitation and the responses is approximately the same, the $H_v$-estimator is used for the FRF estimation. This estimator can handle noise on the input and output simultaneously~\cite{Brandt2023}.

\begin{figure}[H]
	\centering
	\scriptsize
	\includegraphics{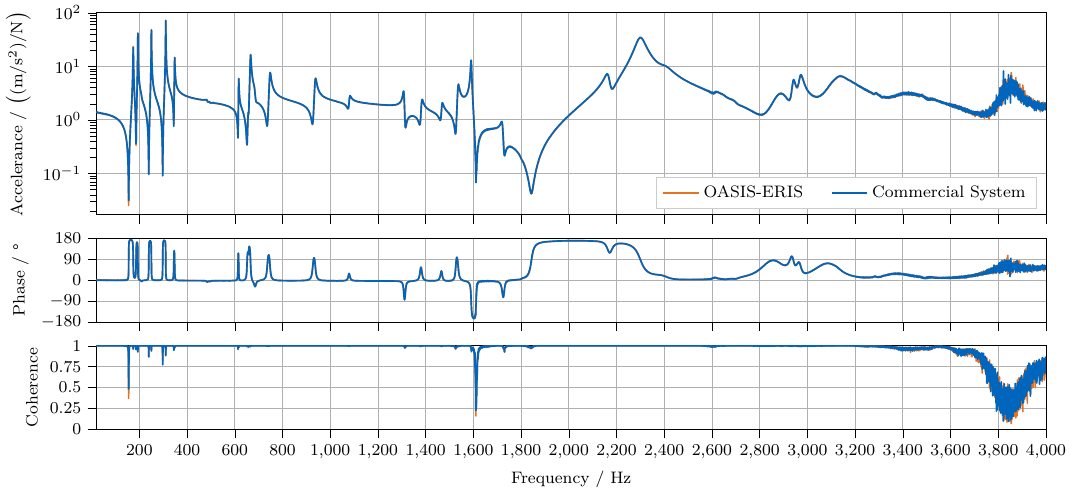}
	\caption{Driving-point FRF of the impedance sensor estimated using the \textit{pyFRF}~\cite{pyFRF} $H_v$-estimator implementation.}
	\label{fig:frf_imp}
\end{figure}

The driving-point FRF of the impedance sensor for the measured data from \textit{OASIS-ERIS} and the \textit{Müller-BBM PAK MKII} system is shown in \cref{fig:frf_imp}. While there are some very minor differences in some of the resonances and anti-resonances, for instance, in the first anti-resonance, the FRFs' magnitude from \textit{OASIS-ERIS} and the commercial system match very closely. The same is true for the phase and coherence plots.

\vspace{-0.4cm}

\begin{figure}[H]
	\centering
	\scriptsize
	\includegraphics{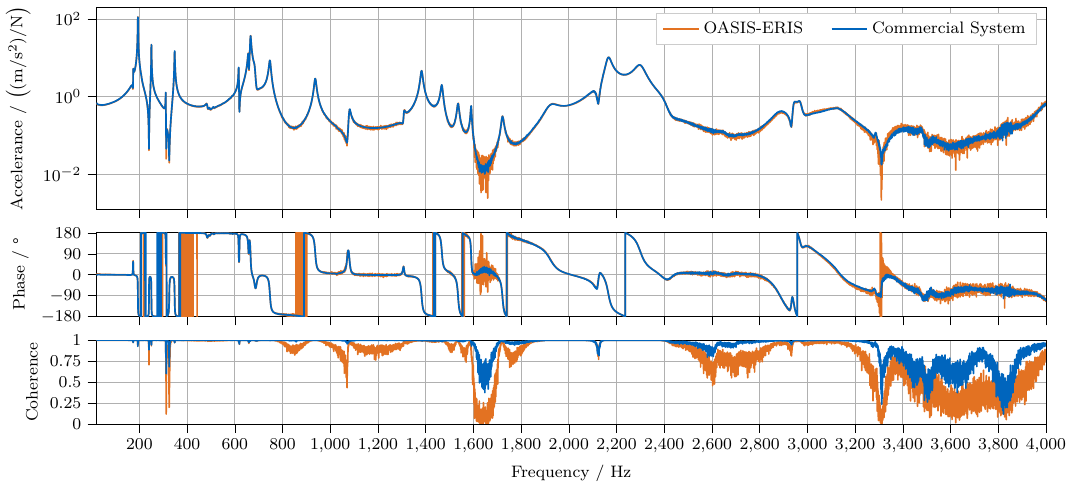}
	\caption{Cross-point FRF between the impedance sensor force and the acceleration response of Sensor 11 in $z$-direction. The FRFs of both systems are estimated using the \textit{pyFRF}~\cite{pyFRF} $H_v$-estimator implementation.}
	\label{fig:frf_s11z}
\end{figure}

A worse performance of \textit{OASIS-ERIS} than the commercial system is observed for sensor degrees of freedom (dofs) with lower signal levels. This is, for instance, the case for the $z$-direction of Sensor 11; see also \cref{fig:frf_s11z}. Especially in the anti-resonance region around \SI{1.6}{\kilo\hertz}, a higher noise level is observed in the FRFs from \textit{OASIS-ERIS}. Also in other areas, where the FRF magnitude is lower, a higher noise level is visible. In these areas, a drop in the coherence for \textit{OASIS-ERIS} is visible.

This shows that \textit{OASIS-ERIS} can match the performance of the commercial system if the signal level is sufficiently high. At lower signal levels, the commercial system proves superior and less susceptible to noise.

\vspace{-0.4cm}

\subsubsection{Experimental modal analysis results}

Lastly, modal parameters are extracted from the FRFs with the modal identification implemented in \textit{pyFBS}~\cite{pyFBS}, using a combination of the poly-reference Least-Squares Complex Frequency (pLSCF) and Least-Squares Frequency Domain (LSFD) method; see~\cite{pyFBSLSCF} for more details. The identified eigenfrequencies and modal parameters in the range from 20 to \SI{2000}{\hertz} are summarized in \cref{tab:EMA_pyFBS}.

As can be seen, there is barely any difference in the identified eigenfrequencies between \textit{OASIS-ERIS} and the commercial system. The biggest relative difference of -\SI{0.06}{\percent} is observed for the first mode; however, in absolute numbers, the difference is only \SI{0.1}{\hertz}, which matches the frequency resolution of the FRFs. In general, the difference in the identified eigenfrequencies is minuscule and within the expected experimental variations. The same is true for the identified damping ratios, where the largest absolute difference of \SI{0.03}{\percent} is observed for the first and tenth modes.

\begin{table}[H]
	\centering
	\scriptsize
	\caption{Summary of identified modal parameters using \textit{pyFBS} modal identification~\cite{pyFBSLSCF}.}
	\label{tab:EMA_pyFBS}
	\begin{tabular}{c|c|c|c|c|c|c}
		& \multicolumn{3}{c|}{\textbf{{Eigenfrequencies $f_i$}}} & \multicolumn{3}{c}{\textbf{Damping ratios $\vartheta_i$}} \\\cline{2-7}
		& \multirow{2}{*}{PAK} & \multirow{2}{*}{OASIS} & Relative & \multirow{2}{*}{PAK} & \multirow{2}{*}{OASIS} & Absolute \\
		& & & Difference & & & Difference \\\hline
		Mode 1 & \SI{174.4}{\hertz} & \SI{174.3}{\hertz} & \SI{-0.06}{\percent} & \SI{0.43}{\percent} & \SI{0.40}{\percent} & -\SI{0.03}{\percent} \\\hline
		Mode 2 & \SI{193.9}{\hertz} & \SI{193.9}{\hertz} & \SI{0.00}{\percent} & \SI{0.26}{\percent} & \SI{0.25}{\percent} & -\SI{0.01}{\percent} \\\hline
		Mode 3 & \SI{250.5}{\hertz} & \SI{250.5}{\hertz} & \SI{0.00}{\percent} & \SI{0.19}{\percent} & \SI{0.19}{\percent} & \SI{0.00}{\percent} \\\hline
		Mode 4 & \SI{310.7}{\hertz} & \SI{310.7}{\hertz} & \SI{0.00}{\percent} & \SI{0.14}{\percent} & \SI{0.14}{\percent} & \SI{0.00}{\percent} \\\hline
		Mode 5 & \SI{348.0}{\hertz} & \SI{348.0}{\hertz} & \SI{0.00}{\percent} & \SI{0.24}{\percent} & \SI{0.25}{\percent} & +\SI{0.01}{\percent} \\\hline
		Mode 6 & \SI{616.5}{\hertz} & \SI{616.5}{\hertz} & \SI{0.00}{\percent} & \SI{0.12}{\percent} & \SI{0.12}{\percent} & \SI{0.00}{\percent} \\\hline
		Mode 7 & \SI{663.6}{\hertz} & \SI{663.4}{\hertz} & \SI{-0.03}{\percent} & \SI{0.28}{\percent} & \SI{0.29}{\percent} & +\SI{0.01}{\percent} \\\hline
		Mode 8 & \SI{747.3}{\hertz} & \SI{747.3}{\hertz} & \SI{0.00}{\percent} & \SI{0.49}{\percent} & \SI{0.49}{\percent} & \SI{0.00}{\percent} \\\hline
		Mode 9 & \SI{937.2}{\hertz} & \SI{937.3}{\hertz} & +\SI{0.01}{\percent} & \SI{0.46}{\percent} & \SI{0.45}{\percent} & -\SI{0.01}{\percent} \\\hline
		Mode 10 & \SI{1080.6}{\hertz} & \SI{1080.8}{\hertz} & +\SI{0.02}{\percent} & \SI{0.41}{\percent} & \SI{0.44}{\percent} & +\SI{0.03}{\percent} \\\hline
		Mode 11 & \SI{1309.0}{\hertz} & \SI{1309.1}{\hertz} & +\SI{0.01}{\percent} & \SI{0.24}{\percent} & \SI{0.24}{\percent} & \SI{0.00}{\percent} \\\hline
		Mode 12 & \SI{1382.9}{\hertz} & \SI{1383.0}{\hertz} & +\SI{0.01}{\percent} & \SI{0.32}{\percent} & \SI{0.32}{\percent} & \SI{0.00}{\percent} \\\hline
		Mode 13 & \SI{1467.9}{\hertz} & \SI{1468.0}{\hertz} & +\SI{0.01}{\percent} & \SI{0.29}{\percent} & \SI{0.29}{\percent} & \SI{0.00}{\percent} \\\hline
		Mode 14 & \SI{1535.9}{\hertz} & \SI{1535.9}{\hertz} & \SI{0.00}{\percent} & \SI{0.27}{\percent} & \SI{0.27}{\percent} & \SI{0.00}{\percent} \\\hline
		Mode 15 & \SI{1590.8}{\hertz} & \SI{1590.9}{\hertz} & +\SI{0.01}{\percent} & \SI{0.13}{\percent} & \SI{0.13}{\percent} & \SI{0.00}{\percent} \\\hline
		Mode 16 & \SI{1722.0}{\hertz} & \SI{1722.2}{\hertz} & +\SI{0.01}{\percent} & \SI{0.29}{\percent} & \SI{0.29}{\percent} & \SI{0.00}{\percent}
	\end{tabular}
\end{table}\vspace{-0.4cm}

\begin{figure}[H]
	\centering
	\scriptsize
	\input{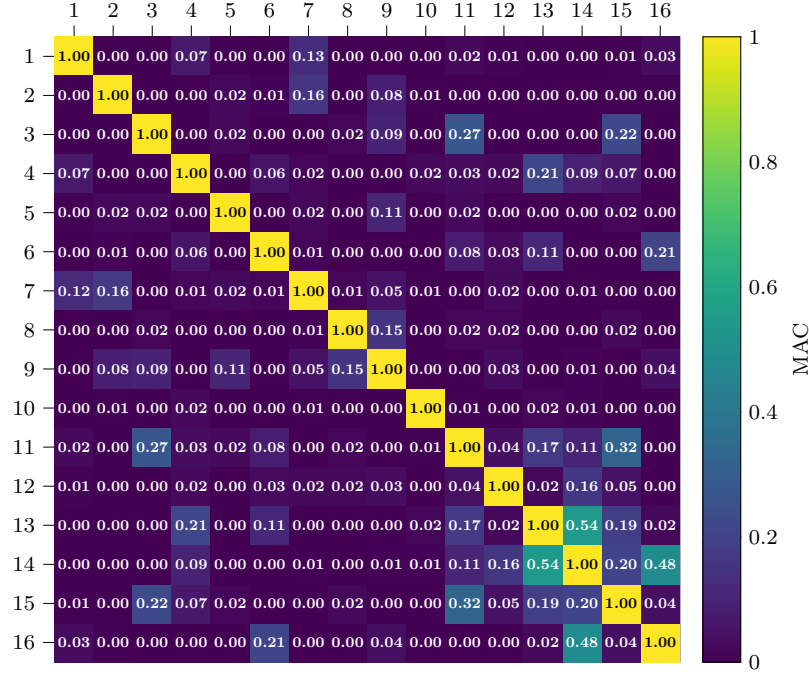}
	\caption{Cross-Modal Assurance Criterion between the mode shapes identified from the measurements of the commercial system and \textit{OASIS-ERIS}.}
	\label{fig:cross_mac}
\end{figure}

A comparison of the Cross-Modal Assurance Criterion (Cross-MAC) between the two identified mode sets, see also \cref{fig:cross_mac}, shows that the identified mode shapes are identical within the accuracy of the calculation. The non-zero off-diagonal values of the Cross-MAC match the value of the Auto-Modal Assurance Criterion, and are not a result of different mode shapes.

\vspace{-0.4cm}

\subsection{Summary and conclusion of \textit{OASIS-ERIS} validation}

The signal generator synchronicity test showed that inter-device synchronization with a maximum offset of one sample, or \SI{10}{\micro\second}, is possible. Further, when compared with the commercial measurement system, it was shown that the performance of \textit{OASIS-ERIS} closely matches that of the commercial system for FRF estimation, provided the signal level is sufficiently high. For the experimental modal analysis campaign, no meaningful differences were observed in the identified modal parameters. In particular, the exact match of the identified mode shapes indicates that inter-device synchronization is successful, since this requires responses from all sensors connected to different \textit{OASIS-ERIS} boards.

With the added inter-device synchronization, the improved maximum sample rate of \SI{100}{\kilo\hertz}, as well as the new software stack offering a graphical user interface, a terminal user interface, and an API for custom scripting, \textit{OASIS-ERIS} is a substantial step toward a fully featured acquisition system for day-to-day lab use.

% !TeX spellcheck = en_US

\noindent
\textbf{CRediT author statement}\\
\noindent
\textbf{O.~M.~Zobel:} Conceptualization; Data curation; Formal analysis; Investigation; Methodology; Project administration; Software; Validation; Visualization; Writing – original draft; Writing – review \& editing.
\textbf{J.~Maierhofer:} Conceptualization; Formal analysis; Investigation; Methodology; Software; Validation; Writing – review \& editing.
\textbf{D.~J.~Rixen:} Resources; Supervision; Writing – review \& editing.
% !TeX spellcheck = en_US

\noindent
\textbf{Acknowledgements}\\
This research did not receive any specific grant from funding agencies in the public, commercial, or not-for-profit sectors.\\
\newpage
\noindent\textbf{References}
\printbibliography[heading=none]

\newpage
\appendix
% !TeX spellcheck = en_US

\section{Command reference}\label{sec:CommandRef}

Note: Here, the command reference at the point of publicationis included; the current version can be found at: \url{https://gitlab.com/oasis-acquisition/oasis-commandreference}

\begin{center}
	\vfill
	\fbox{\includegraphics[scale=0.72,page=1]{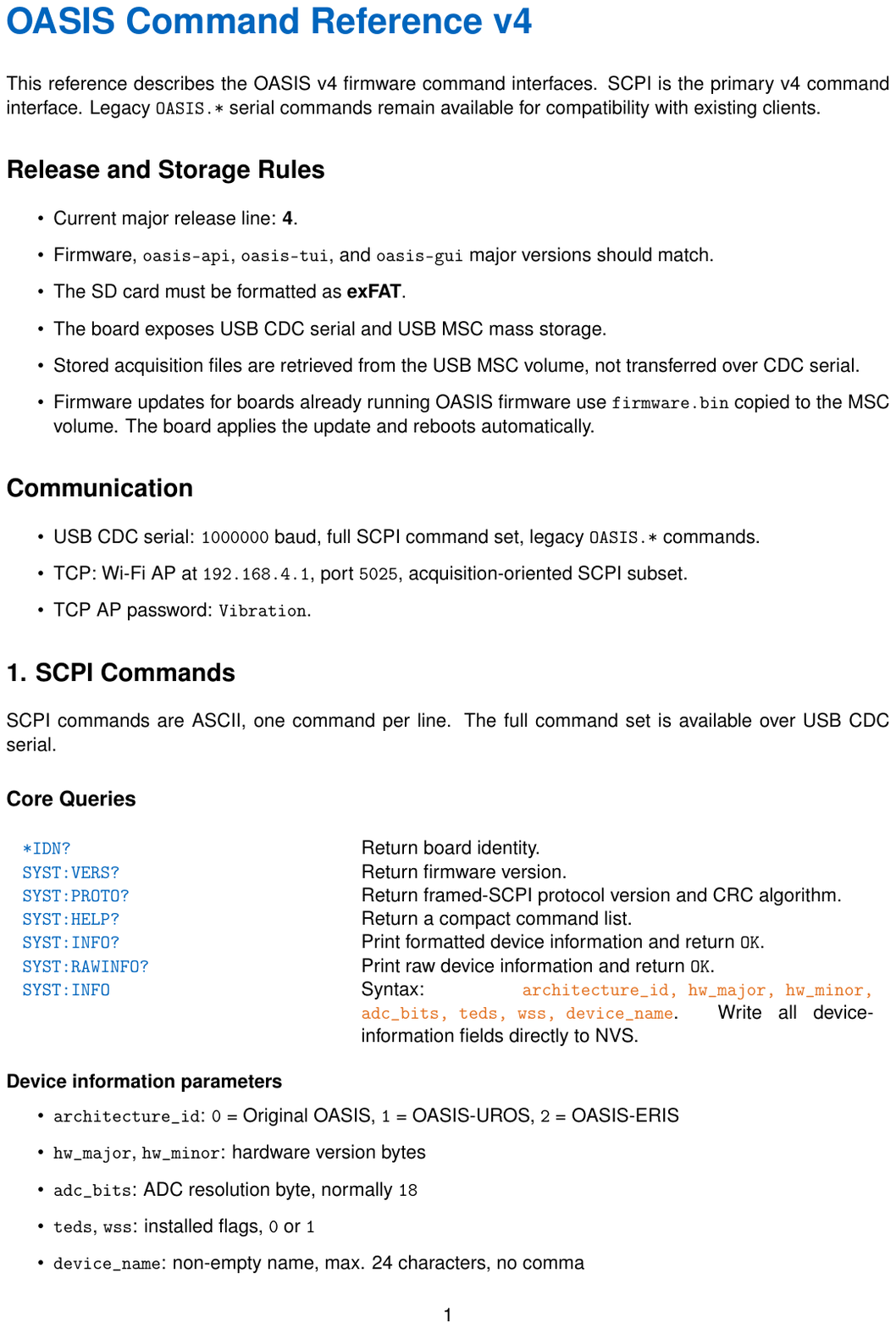}}
	\vfill
\end{center}
\newpage
$ $
\begin{center}
	\vfill
	\fbox{\includegraphics[scale=0.72,page=2]{OASIS-Command-Reference.pdf}}
	\vfill
\end{center}
\newpage
$ $
\begin{center}
	\vfill
	\fbox{\includegraphics[scale=0.72,page=3]{OASIS-Command-Reference.pdf}}
	\vfill
\end{center}
\newpage
$ $
\begin{center}
	\vfill
	\fbox{\includegraphics[scale=0.72,page=4]{OASIS-Command-Reference.pdf}}
	\vfill
\end{center}
\newpage

\end{document}